# Statistical Multiplexing and Traffic Shaping Games for Network Slicing

Jiaxiao Zheng*, Pablo Caballero*, Gustavo de Veciana*, Seung Jun Baek[†] and Albert Banchs[‡]
*The University of Texas at Austin, TX
[†]Korea University, Korea
[‡]University Carlos III of Madrid & IMDEA Networks Institute, Spain
contact: gustavo@ece.utexas.edu

*Abstract*—Next generation wireless architectures are expected to enable slices of shared wireless infrastructure which are customized to specific mobile operators/services. Given infrastructure costs and the stochastic nature of mobile services' spatial loads, it is highly desirable to achieve efficient statistical multiplexing amongst such slices. We study a simple dynamic resource sharing policy which allocates a 'share' of a pool of (distributed) resources to each slice–Share Constrained Proportionally Fair (SCPF). We give a characterization of SCPF's performance gains over static slicing and general processor sharing. We show that higher gains are obtained when a slice's spatial load is more 'imbalanced' than, and/or 'orthogonal' to, the aggregate network load, and that the overall gain across slices is positive. We then address the associated dimensioning problem. Under SCPF, traditional network dimensioning translates to a coupled share dimensioning problem, which characterizes the existence of a feasible share allocation given slices' expected loads and performance requirements. We provide a solution to robust share dimensioning for SCPF-based network slicing. Slices may wish to unilaterally manage their users' performance via admission control which maximizes their carried loads subject to performance requirements. We show this can be modeled as a 'traffic shaping' game with an achievable Nash equilibrium. Under high loads, the equilibrium is explicitly characterized, as are the gains in the carried load under SCPF vs. static slicing. Detailed simulations of a wireless infrastructure supporting multiple slices with heterogeneous mobile loads show the fidelity of our models and range of validity of our high load equilibrium analysis.

## I. Introduction

Next generation wireless systems are expected to embrace SDN/NFV technologies towards realizing slices of shared wireless infrastructure which are customized for specific mobile services, e.g., mobile broadband, media, OTT service providers, and machine-type communications. Customization of network slices may include allocation of (virtualized) resources (communication/computation), per-slice policies, performance monitoring and management, security, accounting, etc. The ability to deploy service specific slices is viewed, not only as a mean to meet the diverse and sometimes stringent demands of emerging services, e.g., vehicular, augmented reality, but also as an approach for infrastructure providers to reduce costs while developing new revenue streams. Resource allocation virtualization in this context is more challenging than for traditional cloud computing. Indeed, rather than drawing on a centralized pool of resources, a network slice requires allocations across a distributed pool of resources, e.g., base stations. The challenge is thus to promote efficient *statistical multiplexing* amongst slices over pools of shared resources.

Network slices can be used to enable the sharing of network resources amongst competing (possibly virtual) operators. Indeed, the sharing of spectrum and infrastructure is viewed as one way of reducing capital/operational costs and is already being considered by standardization bodies, see [1], [2], which have specified architectural and technical requirements, but left the sharing criteria and algorithmic issues open. By aggregating their traffic onto shared resources, it is expected that operators could realize substantial savings, which might justify/enable new shared investments in next generation technologies including 5G, mmWave and massive MIMO.

The focus of this paper is on resource sharing amongst slices supporting stochastic (mobile) loads. A natural approach to sharing is complete partitioning (see, e.g., [16]), which we refer to as *static slicing*, whereby resources are statically partitioned and allocated to slices, according to a service level agreement, irrespective of slices' instantaneous loads. This offers each slice a guaranteed allocation at each base station, and protection from each other's traffic, but, as we will see, poor efficiency. Other approaches include full sharing (where all slices are served on a FCFS basis without resource reservation), general processor sharing [24], which pre-assigns a share to each slice, and allocate resource at each base station proportionally to the shares among the slices which has active users there. Instead, we advocate an alternative approach wherein each slice is pre-assigned a *fixed* share of the pool of resources, and re-distributes its share equally amongst its active customers. In turn, each base station allocates resources to customers in proportion to their shares. We refer to this sharing model as Share Constrained Proportionally Fair (SCPF) resource allocation. By contrast with static slicing, SCPF is *dynamic* (since its resource allocations depend on the network state) but constrained by the network slices' pre-assigned shares (which provides a degree of protection amongst slices).

*Related work.* There is an enormous amount of related work on network resource sharing in the engineering, computer science and economics communities. The standard framework used in the design and analysis of communication networks is utility maximization (see e.g., [28] and references there-in) which has led to the design of several transport and scheduling mechanisms and criteria, e.g., the often considered proportional fair criterion. The SCPF mechanism, described above, should be viewed as a Fisher market where agents (slices), which are share (budget) constrained, bid on network resources, see, e.g., [23], and for applications [5], [10], [15]. The choice to re-distribute a slice's share (budget) equally amongst its users, can be viewed as a network mandated policy, but also emerges naturally as the social optimal, market and Nash equilibrium when slices exhibit (price taking) strategic behavior in optimizing their own utility, see [9].

The novelty of our work lies in considering slice based sharing, under stochastic loads and in particular studying the expected performance resulting from such SCPF-based resource allocations among coupled slices. Other researchers who have considered performance of stochastic networks, e.g., [7], [12], and others, have studied networks where customers are allocated resources (along routes) based on maximizing a sum of customers utilities. These works focus on network stability for 'elastic' customers, e.g., file transfers. Subsequently [8], [27] extended this line of work, to the evaluation of mean file delays, but only under balanced fair resource allocations (as a proxy for proportional fairness). Our focus here is on SCPF-based sharing amongst slices with stochastic loads and on 'inelastic' or 'rate-adaptive' customers, e.g., video, voice, and more generally customers on properly provisioned networks, whose activity on the network can be assumed to be independent of their resource allocations.

Finally there is much ongoing work on developing the network slicing concept, see e.g., [25], [32] and references therein, including development of approaches to network virtualization in RAN architectures, e.g, [11], [20], [26], and SDN-based implementation, e.g., [6]. This paper focuses on devising good slice-based resource sharing criteria to be incorporated into such architectures.

*Contributions of this paper.* This paper makes several contributions centering on a simple and practical resource sharing mechanism: SCPF. First, we consider user performance (bit transmission delay) on slices supporting stochastic loads. In particular we develop expressions for $(i)$ the mean performance seen by a typical user on a network slice; and $(ii)$ the achievable performance gains versus static slicing (SS) and general processor sharing (GPS). We show that when a slice's load is more 'imbalanced' than, and/or 'orthogonal' to, the aggregate network load, one will see higher performance gains. Our analysis provides an insightful picture of the 'geometry' of statistical multiplexing for SCPF-based network slicing. Second, under SCPF, traditional network dimensioning translates to a coupled share dimensioning problem, which addresses whether there exist feasible share allocations given slices' expected loads and performance requirements. We provide a solution to robust share dimensioning for SCPF-based network slicing. Third, we consider decentralized per-slice performance management under SCPF sharing. In particular, we consider admission control aimed at maximizing a slice's carried load subject to a performance constraint. When slices unilaterally optimize their admission control policies, the coupling of their decisions can be viewed as a 'traffic shaping' game, which is shown to have a Nash equilibrium. For a high load regime we explicitly characterize the equilibrium and the associated gains in carried load for SCPF versus static slicing. Finally, we present detailed simulations for a shared distributed infrastructure supporting slices with mobility patterns different than that assumed in the theoretical analysis and more practical SINR model. The results match our analysis well, which further supports our conclusions on gains in both performance and carried loads of SCPF sharing.

## II. SYSTEM MODEL

*A. Network Slices, Resources and Mobile Service Traffic*

We consider a collection of base stations (sectors) $\mathcal{B}$ shared by a set of network slices $\mathcal{V}$, with cardinalities $B$ and $V$ respectively. For example, $\mathcal{V}$ might denote slices supporting different services or (virtual) mobile operators, etc.

We envisage each slice $v$ providing a mobile service in the region served by the base stations $\mathcal{B}$. Each slice supports a stochastic load of users (devices/customers) with an associated mobility/handoff policy. In particular, we assume that exogenous arrivals to slice $v$ at base station $b$ follow a Poisson process with intensity $\gamma_b^v$ and let $\boldsymbol{\gamma}^v$ denote the (column) vector of arrival intensities at each base station associated with slice $v$, i.e. $\boldsymbol{\gamma}^v = (\gamma_b^v : b \in \mathcal{B})$. Each slice $v$ customer at base station $b$ has an independent sojourn time with mean $\mu_b^v$ after which it is randomly routed to another base station or exits the system. As explained below we assume that such mobility patterns do not depend on the resources allocated to users. We let $\boldsymbol{Q}^v = (q_{i,j}^v : i, j \in \mathcal{B})$ denote a slice-dependent routing matrix where $q_{i,j}^v$ is the probability a slice $v$ customer moves from base station $i$ to $j$ and $1 - \sum_{j \in \mathcal{B}} q_{i,j}^v$ is the probability it exits the system. This model induces an overall traffic intensity for slice $v$ across base stations satisfying flow conservation equations: for all $b \in \mathcal{B}$ we have

$$\kappa_b^v = \gamma_b^v + \sum_{a \in \mathcal{B}} \kappa_a^v q_{a,b}^v,$$

where $\kappa_b^v$ is the traffic intensity of slice $v$ on base station $b$. Accounting for users' sojourn times, the mean offered load of slice $v$ on base station $b$ is $\rho_b^v = \kappa_b^v \mu_b^v$, and $\boldsymbol{\rho}^v \triangleq (\rho_b^v : b \in \mathcal{B})$ captures its system load distribution. Letting $\boldsymbol{\mu}^v = (\mu_b^v : b \in \mathcal{B})$, the flow conservation equations can be rewritten in matrix form as:

$$\boldsymbol{\rho}^v = \text{diag}(\boldsymbol{\mu}^v)(\boldsymbol{I} - (\boldsymbol{Q}^v)^T)^{-1}\boldsymbol{\gamma}^v. \quad (1)$$

If $\boldsymbol{Q}^v$ is irreducible, $\boldsymbol{I} - (\boldsymbol{Q}^v)^T$ is irreducibly diagonally dominant thus always invertible. Otherwise, we can always find a permutation matrix of $\mathcal{B}$, say $\boldsymbol{P}$ to make:

$$\boldsymbol{P}^T(\boldsymbol{I} - (\boldsymbol{Q}^v)^T)\boldsymbol{P} = \begin{bmatrix} \boldsymbol{A}_1 & \boldsymbol{B}_{1,2} & \cdots \\ & \ddots & \vdots \\ & & \boldsymbol{A}_K \end{bmatrix},$$

where $K$ is the number of irreducible classes. Moreover, at least one base station of each irreducible class has a nonzero exiting probability, thus $\boldsymbol{A}_K$ must be invertible. Then the invertibility of $\boldsymbol{I} - (\boldsymbol{Q}^v)^T$ follows.

This model corresponds to a multi-class network of $M/GI/\infty$ queues (base stations), where each slice corresponds to a class of customers, see, e.g., [19]. Such networks are known to have a *product-form stationary* distribution, i.e., the numbers of customers on slice $v$ at base station $b$, denoted by $N_b^v$, are mutually independent and $N_b^v \sim \text{Poisson}(\rho_b^v)$. Since the sum of independent Poisson random variables is again Poisson, the total number of customers on slice $v$ is such that $N^v = \sum_{b \in \mathcal{B}} N_b^v \sim \text{Poisson}(\rho^v)$ where $\rho^v \triangleq \sum_{b \in \mathcal{B}} \rho_b^v$.

Our network model for the numbers of customers and mobility across base stations, assumes that customer sojourn/activity/mobility are independent of the network state and of the resources a customer is allocated. This is reasonable for properly engineered slices where the performance a customer sees does not impact its activity, e.g., *inelastic* or *rate adaptive* applications seeing acceptable performance. This covers a wide range of applications including voice, video streaming, IoT monitoring, real-time control, and even, to some degree, elastic web browsing sessions where users are peak rate constrained and this constraint typically dictates their performance.

There are several natural generalizations to this model including class-based routing and user sessions (e.g. web browsing) which are not always active at the base stations they visit, see, e.g., [19].

### B. Network Slice Resource Sharing

In the sequel we consider a setting where the resources allocated to a slice's customers depend on the overall network state, i.e., number of customers each slice has on each base station, corresponding to the stochastic process described in Section II-A. Let us consider a *snapshot* of the system's state and let $\mathcal{U}_b^v, \mathcal{U}_b, \mathcal{U}^v$, and $\mathcal{U}$ denote sets of active customers on slice $v$ at base station $b$, at base station $b$, on slice $v$, and on the overall network, respectively. Thus, the cardinalities of these sets correspond to a realization of the system 'state', i.e., $|\mathcal{U}_b^v| = n_b^v$ and $|\mathcal{U}^v| = n^v$, where in a stationary regime $n^v$ and $n_b^v$ are realizations of Poisson random variables $N^v$ and $N_b^v$, respectively.

Each base station $b$ is modeled as a finite resource shared by its associated users $\mathcal{U}_b$. A customer $u \in \mathcal{U}_b$ can be allocated a fraction $f_u \in [0,1]$ of that resource, e.g., of resource blocks in a given LTE frame, or allocated the resource for a fraction of time, where $\sum_{u \in \mathcal{U}_b} f_u = 1$. We shall neglect quantization effects. The transmission rate to customer $u$, denoted by $r_u$, is then given by $r_u = f_u c_u$ where $c_u$ denotes the current peak rate for that user. To model customer heterogeneity across slices/base stations we shall assume $c_u$ for a typical customer on slice $v$ at base station $b$ is an independent realization of a random variable, denoted by $C_b^v$, whose distribution may depend on the slice, since slices may support different types of customer devices (e.g., car connectivity vs. mobile phone) and depend on the base station, since typical slice $v$ users may have different spatial distributions with respect to base station $b$ or see different levels of interference.

Below we consider three resource allocation schemes; the first two are used as benchmarks, while the third is the one under study in this paper. For all we assume each slice is allocated a 'share' of the network resources $s^v, v \in \mathcal{V}$ such that $s^v > 0$ and $\sum_{v \in \mathcal{V}} s^v = 1$.

**Definition 1.** *Static Slicing (SS): Under SS, slice $v$ is allocated a fixed fraction $s^v$ of each base station $b$'s resources, and each customer $u \in \mathcal{U}_b^v$ gets an equal share, i.e., $1/n_b^v$, of the slice $v$'s resources at base station $b$. Thus the users transmission rate $r_u^{SS}$ is given by*

$$r_u^{SS} = \frac{s^v}{n_b^v} c_u.$$

**Definition 2.** *General Processor Sharing (GPS):* [24] *Under GPS, each active slice $v$ at base station $b$ such that $n_b^v > 0$ is allocated a fraction of the base station $b$'s resources proportionally to its share $s^v$. Thus a user $u \in \mathcal{U}_b^v$ sees a transmission rate $r_u^{GPS}$ given by*

$$r_u^{GPS} = \frac{s^v}{n_b^v \sum_{v' \in \mathcal{V}} s^{v'} \mathbf{1}_{\{n_b^{v'} > 0\}}} c_u. \quad (2)$$

**Definition 3.** *Share Constrained Proportionally Fair (SCPF): Under SCPF each slice re-distributes its share of the overall network resources equally amongst its active customers, which thus get a weight (sub-share) $w_u = \frac{s^v}{n_v}$ for $u \in \mathcal{U}^v, \forall v \in \mathcal{V}$. In turn, each base station allocates resources to customers in proportion to their weights. So a user $u \in \mathcal{U}_b^v$ gets a transmission rate $r_u^{SCPF}$ given by*

$$r_u^{SCPF} = \frac{w_u}{\sum_{u' \in \mathcal{U}_b} w_{u'}} c_u = \frac{\frac{s_v}{n^v}}{\sum_{v' \in \mathcal{V}} \frac{n_b^{v'} s_{v'}}{n^{v'}}} c_u. \quad (3)$$

A simple example illustrating the differences among three schemes is exhibited in Table I. Suppose there are two base stations, i.e., $\mathcal{B} = \{b_1, b_2\}$, and two slices $\mathcal{V} = \{1, 2\}$ each with an equal share of the network resource. Consider a snapshot of the system where Users $u_1, u_2$ are on Slice 1 and $u_3, u_4$ are on Slice 2. Also, $u_1, u_2$, and $u_3$ are at base station $b_1$ and $u_4$ is at base station $b_2$. Let us assume for simplicity that $c_u = 1, \forall u \in \mathcal{U}$. In this case, under SS at $b_1$ the two users on Slice 1 need to share $\frac{1}{2}$ of the resource while $u_3$ on Slice 2 is allocated the other $\frac{1}{2}$, while at $b_2$, half of the resource is wasted due to the absence of active users on Slice 1. By contrast, GPS utilizes all resources at $b_2$ by allocating all of them to $u_4$, and it makes the same allocation as SS at $b_1$. Under SCPF, because each user is allocated the same weight $\frac{1}{4}$, at $b_1$, three users are allocated the same rate $\frac{1}{3}$ and at $b_2$ all bandwidth is given to $u_4$. This example shows how SCPF achieves better network-wide fairness than GPS and SS, while ensuring that resources are not wasted.

Indeed, under SCPF the *overall* fraction of resources slice $v$ is allocated at base station $b$ is proportional to $\frac{n_b^v}{n^v} s^v$, i.e., its *share* times its *relative* number of users at the base station. This provides a degree of *elasticity* to variations in the slice's spatial loads. However, if a slice has a large number of customers, its customers' weights are proportionally decreased, which protects other slices from such overloads. Note that SCPF requires minimal information exchanges among base stations and is straightforward to implement, e.g., using SDN-like framework.

TABLE I: Rate allocation under different schemes

| | User association | | Rate allocation | | |
|---|---|---|---|---|---|
| | Slice 1 | Slice 2 | SS | GPS | SCPF |
| $b_1$ | $u_1$ | | $\frac{1}{4}$ | $\frac{1}{4}$ | $\frac{1}{3}$ |
| | $u_2$ | | $\frac{1}{4}$ | $\frac{1}{4}$ | $\frac{1}{3}$ |
| | | $u_3$ | $\frac{1}{2}$ | $\frac{1}{2}$ | $\frac{1}{3}$ |
| $b_2$ | | $u_4$ | $\frac{1}{2}$ | 1 | 1 |

## III. PERFORMANCE EVALUATION

In this section we study the expected performance seen by a slice's typical customer. Given our focus on inelastic/rate adaptive traffic and tractability, we choose our customer performance metric as the reciprocal transmission rate, referred to as the *Bit Transmission Delay (BTD)*, see, e.g., [30]. This corresponds to the time taken to transmit a 'bit', so lower BTDs indicate higher rates and thus better performances. BTD is a high-level metric capturing the instantaneous QoS perceived by a user, e.g., short packet transmission delays are roughly proportional to the BTD. By guaranteeing a good BTD we can guarantee that the user perceived QoS is acceptable all the time, instead of in an average sense. Alternatively, the negative of the BTD can be viewed as a concave utility function of the rate, which in the literature (see, e.g., [22]) was referred to as the *potential delay* utility. Concave utility functions tend to favor allocations that exhibit reduced variability in a stochastic setting. Given the stochastic loads on the network, we shall evaluate the average BTD seen by a typical (i.e., randomly selected) customer on a slice, i.e., averaged over the stationary distribution of the network state and transmission capacity seen by typical users, e.g., $C_b^v$, at each base station. Such averages naturally place higher weights on congested base stations, where a slice may have more users, best reflecting the overall performance customers will see.

### A. Analysis of BTD Performance

Consider a *typical* customer on slice $v$ and let $\mathbb{E}^v$ denote the expectation of the system state as seen by such a customer, i.e., under the Palm distribution [4]. For SCPF, we let $R^v$ be a random variable denoting the rate of a typical customer on slice $v$, and $R_b^v$ that of such customer on slice $v$ at base station $b$. Similarly, let $R^{v,SS}$, $R_b^{v,SS}$, $R^{v,GPS}$, and $R_b^{v,GPS}$ denote these quantities under SS and GPS, respectively. Thus, under SCPF the average BTD for a typical slice $v$ customer is given by $\mathbb{E}^v[\frac{1}{R^v}]$. The next result characterizes the mean BTD under SCPF, SS, and GPS under our traffic model. We introduce some further notation:

- Overall average number of users on slice $v$: $\rho^v = \mathbb{E}[N^v]$.
- Load distribution of slice $v$: $\boldsymbol{\rho}^v \triangleq (\rho_b^v : b \in \mathcal{B})$, where $\rho_b^v = \mathbb{E}[N_b^v]$.
- Relative load distribution of slice $v$: $\tilde{\boldsymbol{\rho}}^v \triangleq (\tilde{\rho}_b^v : b \in \mathcal{B})$, where $\tilde{\rho}_b^v = \frac{\rho_b^v}{\rho^v}$.
- Overall share weighted relative load distribution: $\tilde{\boldsymbol{g}} \triangleq (\tilde{g}_b : b \in \mathcal{B})$, where $\tilde{g}_b = \sum_{v \in \mathcal{V}} s^v \tilde{\rho}_b^v$.
- Active share weighted relative load distribution: $\tilde{\boldsymbol{g}}' \triangleq (\tilde{g}_b' : b \in \mathcal{B})$, where $\tilde{g}_b' = \sum_{v \in \mathcal{V}} s^v (1 - e^{-\rho^v}) \tilde{\rho}_b^v$. If a slice is inactive, i.e., not having any customer in the network, then its share should be voided. $(1 - e^{-\rho^v})$ is the probability that slice $v$ has at least 1 active user in the network.
- Vector of idle shares seen by a typical user on slice $v$: $\bar{\boldsymbol{s}}^v \triangleq (\bar{s}_b^v : b \in \mathcal{B})$, where $\bar{s}_b^v = \mathbb{E}^v \left[ \sum_{v' \neq v} s^{v'} \mathbf{1}_{\{N_b^{v'}=0\}} \right] = \sum_{v' \neq v} s^{v'} e^{-\rho_b^{v'}}$ represents how much share is voided seen by slice $v$, due to the absence of active users at base station $b$. Also we define $\boldsymbol{S}^v = \text{diag}(\bar{\boldsymbol{s}}^v)$.
- Mean reciprocal capacity of slice $v$: $\boldsymbol{\delta}^v \triangleq (\delta_b^v : b \in \mathcal{B})$, where $\delta_b^v = \mathbb{E}^v[\frac{1}{C_b^v}]$. Also we define $\boldsymbol{\Delta}_v = \text{diag}(\boldsymbol{\delta}^v)$.

We use $\langle \boldsymbol{x}_1, \boldsymbol{x}_2 \rangle_{\boldsymbol{M}} \triangleq \boldsymbol{x}_1^T \boldsymbol{M} \boldsymbol{x}_2$ to denote the weighted inner product of vectors, where $\boldsymbol{M}$ is a diagonal matrix. Also, we use $\|\boldsymbol{x}\|_{\boldsymbol{M}} \triangleq \sqrt{\boldsymbol{x}^T \boldsymbol{M} \boldsymbol{x}}$ to denote the weighted norm of a vector, where $\boldsymbol{M}$ is a diagonal matrix. In both cases, when $\boldsymbol{M}$ is the identity matrix $\boldsymbol{I}$ we simply omit it. In addition, $\|\boldsymbol{x}\|_2$ and $\|\boldsymbol{x}\|_1$ denote the L2-norm and L1-norm of $\boldsymbol{x}$, respectively.

**Theorem 1.** *For network slicing based on SCPF, the mean BTD for a typical customer on slice $v$ is given by*

$$\mathbb{E}^v \left[ \frac{1}{R^v} \right] = \sum_{b \in \mathcal{B}} \tilde{\rho}_b^v \delta_b^v \left( 1 - \tilde{\rho}_b^v + (\rho^v + 1) \left( \frac{\tilde{g}_b'}{s^v} + e^{-\rho^v} \tilde{\rho}_b^v \right) \right). \quad (4)$$

*If $(\tilde{\boldsymbol{\rho}}^v : v \in \mathcal{V})$ are fixed, and $(\rho^v : v \in \mathcal{V})$ are large, then the mean BTD has following asymptotic form:*

$$\mathbb{E}^v \left[ \frac{1}{R^v} \right] \approx \frac{\rho^v}{s^v} \langle \tilde{\boldsymbol{\rho}}^v, \tilde{\boldsymbol{g}} \rangle_{\boldsymbol{\Delta}_v} + O(1) \quad (5)$$

*For network slicing based on SS, the mean BTD for a typical customer on slice $v$ is given by*

$$\mathbb{E}^v \left[ \frac{1}{R^{v,SS}} \right] = \sum_{b \in \mathcal{B}} \tilde{\rho}_b^v \delta_b^v \left( \frac{\rho_b^v + 1}{s^v} \right). \quad (6)$$

*For network slicing based on GPS, the mean BTD for a typical customer on slice $v$ is given by*

$$\mathbb{E}^v \left[ \frac{1}{R^{v,GPS}} \right] = \sum_{b \in \mathcal{B}} \tilde{\rho}_b^v \delta_b^v \left( \frac{\rho_b^v + 1}{s^v} \right) (1 - \bar{s}_b^v). \quad (7)$$

Please see appendix for the detailed proof. BTD under all 3 schemes increases with the overall load $\rho^v$ and decreases with the share $s^v$ when $(\tilde{\boldsymbol{\rho}}^v : v \in \mathcal{V})$ are fixed. Their dependencies on relative loads $(\tilde{\boldsymbol{\rho}}^v : v \in \mathcal{V})$ are different, implying that they exploit statistical multiplexing differently.

*B. Analysis of Gain*

Using the results in Theorem 1 one can evaluate the gains in the mean BTD for a typical slice $v$ user under SCPF vs. SS, defined as,

$$G_v^{SS} \triangleq \frac{\mathbb{E}^v\left[\frac{1}{R^{v,SS}}\right]}{\mathbb{E}^v\left[\frac{1}{R^v}\right]}.$$

In general, one would expect $G_v^{SS} \geq 1$ since under SCPF typical users should see higher allocated rates and thus lower BTDs. One can verify that is the case when slices have *uniform* loads across base stations but the general case is more subtle. Similarly, we define the gain of SCPF vs. GPS by

$$G_v^{GPS} \triangleq \frac{\mathbb{E}^v\left[\frac{1}{R^{v,GPS}}\right]}{\mathbb{E}^v\left[\frac{1}{R^v}\right]}.$$

By taking the ratio of the mean BTD perceived by a typical customer under SS and that under SCPF given in Theorem 1, we have the following corollary.

**Corollary 1.** *The BTD gain of SCPF over SS for slice $v$ is given by*

$$G_v^{SS} = \frac{\rho^v \|\tilde{\boldsymbol{\rho}}^v\|_{\boldsymbol{\Delta}^v}^2 + \langle \boldsymbol{\delta}^v, \tilde{\boldsymbol{\rho}}^v \rangle}{s^v\langle \boldsymbol{\delta}^v, \tilde{\boldsymbol{\rho}}^v\rangle - s^v(1-(\rho^v+1)e^{-\rho^v})\|\tilde{\boldsymbol{\rho}}^v\|_{\boldsymbol{\Delta}^v}^2 + (\rho^v+1)\langle \tilde{\boldsymbol{g}}', \tilde{\boldsymbol{\rho}}^v\rangle_{\boldsymbol{\Delta}^v}} \quad (8)$$

*For fixed relative loads $(\tilde{\boldsymbol{\rho}}^v : v \in \mathcal{V})$, when slice $v$ has a light load, i.e., $\rho^v \to 0$, the gain is greater than 1 and given by:*

$$G_v^{SS,L} = \frac{\langle \boldsymbol{\delta}^v, \tilde{\boldsymbol{\rho}}^v\rangle}{s^v\langle \boldsymbol{\delta}^v, \tilde{\boldsymbol{\rho}}^v\rangle + \langle \tilde{\boldsymbol{g}}', \tilde{\boldsymbol{\rho}}^v\rangle_{\boldsymbol{\Delta}^v}} > 1$$

*Furthermore, $G_v^{SS}$ is a nonincreasing function of $\rho^v$, and if all slices have high overall loads, i.e., $\rho^v \to \infty, \forall v \in \mathcal{V}$, the gain is given by:*

$$G_v^{SS,H} = \frac{\|\tilde{\boldsymbol{\rho}}^v\|_{\boldsymbol{\Delta}^v}^2}{\langle \tilde{\boldsymbol{g}}, \tilde{\boldsymbol{\rho}}^v\rangle_{\boldsymbol{\Delta}^v}}.$$

The result indicates that when the relative loads are fixed, the gain decreases with the overall load $\rho^v$, thus if $G_v^{SS,H} > 1$ SCPF always provides a gain. Let us consider the heavy load gain under the following simplifying assumption.

**Assumption 1.** *Base stations are said to be homogeneous for slice $v$ if for all $b \in \mathcal{B}$: $\mathbb{E}^v\left[\frac{1}{C_b^v}\right] = \delta_v$.*

Assumption 1 only requires the *average* reciprocal capacity a given slices' customer sees across base stations is homogenous. In this case, the BTD gain for slice $v$ under heavy load simplifies to

$$G_v^{SS,H} = \frac{\|\tilde{\boldsymbol{\rho}}^v\|_2}{\|\tilde{\boldsymbol{g}}\|_2} \times \frac{1}{\cos(\theta(\tilde{\boldsymbol{g}}, \tilde{\boldsymbol{\rho}}^v))},$$

where $\theta(\tilde{\boldsymbol{g}}, \tilde{\boldsymbol{\rho}}^v)$ denotes the angle between the slice's relative load and the overall share weighted relative load on the network. A sufficient condition for gains under high loads is that $\|\tilde{\boldsymbol{g}}\|_2 \leq \|\tilde{\boldsymbol{\rho}}^v\|_2$. Since $\|\tilde{\boldsymbol{g}}\|_1 = \|\tilde{\boldsymbol{\rho}}^v\|_1 = 1$, this follows when the overall share weighted relative load on the network is more balanced than that of slice $v$. One would typically expect aggregated traffic to be more balanced than that of individual slices. This condition is fairly weak, i.e., it does not depend on where the loads are placed, but on how balanced they are. The corollary also suggests that gains are higher when $\cos(\theta(\tilde{\boldsymbol{g}}, \tilde{\boldsymbol{\rho}}^v))$ is smaller. In other words, a slice with imbalanced relative loads whose relative load distribution is 'orthogonal' to the shared weighted aggregate traffic, i.e., $\cos(\theta(\tilde{\boldsymbol{g}}, \tilde{\boldsymbol{\rho}}^v)) \approx 0$, will tend to see higher gains. This is due to that SCPF can achieve sharing elasticity by aligning resource allocations with demands, i.e., load distributions. Thus when the load distributions are nearly orthogonal, sharing under SCPF is much better than that under SS, which is completely inelastic. The simulations in Section VI further explore these observations.

Similarly, for the BTD gain of SCPF over GPS, we have following result:

**Corollary 2.** *The BTD gain of SCPF over GPS for slice $v$ is given by*

$$G_v^{GPS} = \frac{\rho^v\left(\|\tilde{\boldsymbol{\rho}}^v\|_{\boldsymbol{\Delta}^v}^2 - \|\tilde{\boldsymbol{\rho}}^v\|_{\boldsymbol{\Delta}^v \boldsymbol{S}^v}^2\right) + \langle \tilde{\boldsymbol{\rho}}^v, \mathbf{1}-\bar{\boldsymbol{s}}^v\rangle_{\boldsymbol{\Delta}^v}}{s^v\langle \boldsymbol{\delta}^v, \tilde{\boldsymbol{\rho}}^v\rangle - s^v(1-(\rho^v+1)e^{-\rho^v})\|\tilde{\boldsymbol{\rho}}^v\|_{\boldsymbol{\Delta}^v}^2 + (\rho^v+1)\langle \tilde{\boldsymbol{g}}', \tilde{\boldsymbol{\rho}}^v\rangle_{\boldsymbol{\Delta}^v}}. \quad (9)$$

*For fixed relative loads $(\tilde{\boldsymbol{\rho}}^v : v \in \mathcal{V})$, and fixed overall loads for other slices $(\rho^{v'} : v' \neq v)$, the gain for slice $v$ under low overall load, $\rho^v \to 0$, is given by:*

$$G_v^{GPS,L} = \frac{\langle \tilde{\boldsymbol{\rho}}^v, \mathbf{1}-\bar{\boldsymbol{s}}^v\rangle_{\boldsymbol{\Delta}^v}}{s^v\langle \boldsymbol{\delta}^v, \tilde{\boldsymbol{\rho}}^v\rangle + \langle \tilde{\boldsymbol{g}}', \tilde{\boldsymbol{\rho}}^v\rangle_{\boldsymbol{\Delta}^v}}.$$

*Furthermore, if all slices have low load $\rho^v \to 0, \forall v \in \mathcal{V}$, then*

$$G_v^{GPS,L} \to 1$$

*Also, if all slices have high loads, i.e., $\rho^v \to \infty, \forall v \in \mathcal{V}$, the BTD gain over GPS for slice $v$ is given by:*

$$G_v^{GPS,H} = \frac{\|\tilde{\boldsymbol{\rho}}^v\|_{\boldsymbol{\Delta}^v}^2 - \|\tilde{\boldsymbol{\rho}}^v\|_{\boldsymbol{\Delta}^v \boldsymbol{S}^v}^2}{\langle \tilde{\boldsymbol{g}}, \tilde{\boldsymbol{\rho}}^v\rangle_{\boldsymbol{\Delta}^v}}.$$

Please see appendix for detailed proof. Note that when $(\tilde{\boldsymbol{\rho}}^v : v \in \mathcal{V})$ are fixed and $\forall v, b, \tilde{\rho}_b^v > 0$, under heavy load, i.e., $\rho^v \to \infty, \forall v \in \mathcal{V}$, we have $\bar{\boldsymbol{s}}^v \to \mathbf{0}$, thus $\|\tilde{\boldsymbol{\rho}}^v\|_{\boldsymbol{\Delta}^v}^2 - \|\tilde{\boldsymbol{\rho}}^v\|_{\boldsymbol{\Delta}^v \boldsymbol{S}^v}^2 \to \|\tilde{\boldsymbol{\rho}}^v\|_{\boldsymbol{\Delta}^v}^2$, which means GPS obtains a similar performance as SS under heavy load. However, unlike the gain over SS, $G_v^{GPS,L}$ might not be strictly greater than 1 and $G_v^{GPS}$ might not be monotonic in $\rho^v$.

One can observe that, different slices can experience different BTD gains, depending on its share and load distributions. However, to compare the performances of different sharing criteria, a network-wide metric of gain needs to be defined. To be able to compare scenarios with different load distributions and shares, it is of particular interest to consider a metric which is robust against changes of load and share. To devise such a metric, we note that users who are perceiving a low average capacity from their associated base stations, and/or users whose allocated shares are small are expected to experience higher BTDs. Thus to achieve the robustness of the metric, let us define the normalized BTD for a typical user on slice $v$ at base station $b$ under SCPF as

$$\bar{\mathbb{E}}^v\left[\frac{1}{R_b^v}\right] \triangleq \frac{1}{\delta_b^v}\frac{s^v}{\rho^v}\mathbb{E}^v[\frac{1}{R_b^v}], \quad (10)$$

and thus the normalized BTD for a typical user on slice $v$ under SCPF is given by

$$\bar{\mathbb{E}}^v\left[\frac{1}{R^v}\right] = \sum_{b\in\mathcal{B}} \tilde{\rho}_b^v \bar{\mathbb{E}}^v\left[\frac{1}{R_b^v}\right]. \quad (11)$$

Similarly, one can define $\bar{\mathbb{E}}^v\left[\frac{1}{R_b^{v,SS}}\right]$, $\bar{\mathbb{E}}^v\left[\frac{1}{R^{v,SS}}\right]$, and $\bar{\mathbb{E}}^v\left[\frac{1}{R_b^{v,GPS}}\right]$, $\bar{\mathbb{E}}^v\left[\frac{1}{R^{v,GPS}}\right]$. For the overall performance of the system, let us consider the share weighted sum of the normalized BTD since the system should be tuned to put more emphasis on the slices with higher shares, and define the overall weighted BTD gain of SCPF over SS as

$$G_{\text{all}}^{SS} \triangleq \frac{\sum_{v\in\mathcal{V}} s^v \bar{\mathbb{E}}^v\left[\frac{1}{R^{v,SS}}\right]}{\sum_{v\in\mathcal{V}} s^v \bar{\mathbb{E}}^v\left[\frac{1}{R^v}\right]}, \quad (12)$$

and the overall weighted BTD gain of SCPF over GPS as

$$G_{\text{all}}^{GPS} \triangleq \frac{\sum_{v\in\mathcal{V}} s^v \bar{\mathbb{E}}^v\left[\frac{1}{R^{v,GPS}}\right]}{\sum_{v\in\mathcal{V}} s^v \bar{\mathbb{E}}^v\left[\frac{1}{R^v}\right]}, \quad (13)$$

The following results capture the overall weighted BTD gains.

**Corollary 3.** *When $\rho^v \to \infty, \forall v \in \mathcal{V}$, the overall weighted BTD gains of SCPF over SS and GPS under heavy load are given by*

$$G_{\text{all}}^{SS,H} = \frac{\sum_{v\in\mathcal{V}} s^v \|\tilde{\boldsymbol{\rho}}^v\|_2^2}{\sum_{v\in\mathcal{V}} s^v \langle \tilde{\boldsymbol{g}}, \tilde{\boldsymbol{\rho}}^v\rangle}, G_{\text{all}}^{GPS,H} = \frac{\sum_{v\in\mathcal{V}} s^v \|\tilde{\boldsymbol{\rho}}^v\|_{\boldsymbol{I}-\boldsymbol{S}^v}^2}{\sum_{v\in\mathcal{V}} s^v \langle \tilde{\boldsymbol{g}}, \tilde{\boldsymbol{\rho}}^v\rangle},$$
(14)
*and*

$$G_{\text{all}}^{SS,H} \geq 1,\ G_{\text{all}}^{GPS,H} \geq 1.$$

Please see appendix for detailed proof. It is easy to see that if $\tilde{\boldsymbol{\rho}}^v$ are the same for all $v\in\mathcal{V}$, then both $G_{\text{all}}^{SS,H}$ and $G_{\text{all}}^{GPS,H}$ are 1 when the loads are heavy. By contrast, if the relative loads of different slices are (approximately) all orthogonal, i.e., $\langle \tilde{\boldsymbol{\rho}}^v, \tilde{\boldsymbol{\rho}}^{v'}\rangle \cong 0, v\neq v'$ and each slice has the same share $s^v = \frac{1}{V}, \forall v\in\mathcal{V}$, the overall gain can be as high as $V$.

## IV. SHARE DIMENSIONING UNDER SCPF

In practice each slice $v$ may wish to provide service guarantees to its customers, i.e., ensure that the mean BTD does not exceed a performance target $d_v$. Below we investigate how to dimension network shares to support slice loads subject to such mean BTD requirements.

Henceforth we shall assume the following assumption is in effect.

**Assumption 2.** *The network is said to see high overall slice loads, if for all $v\in\mathcal{V}$ we have $\rho^v \gg 1$.*

Consider a network supporting the traffic loads of a *single* slice, say $v$, so $s^v = 1$ and $\tilde{\boldsymbol{g}} = \tilde{\boldsymbol{\rho}}^v$. Note that $\langle \tilde{\boldsymbol{\rho}}^v, \boldsymbol{\delta}^v\rangle$ is the minimum average BTD achievable across the network when a slice gets *all* the base station resources, so a target requirement satisfies $d_v > \langle \tilde{\boldsymbol{\rho}}^v, \boldsymbol{\delta}^v\rangle$. For slice $v$ to meet a mean BTD constraint $d_v$, it follows from Eq. (5) that:

$$\rho^v \leq l(d_v, \tilde{\boldsymbol{\rho}}^v, \boldsymbol{\delta}^v) \triangleq \frac{d_v - \langle \tilde{\boldsymbol{\rho}}^v, \boldsymbol{\delta}^v\rangle}{\|\tilde{\boldsymbol{\rho}}^v\|_{\boldsymbol{\Delta}^v}^2}.$$

We can interpret $l(d_v, \tilde{\boldsymbol{\rho}}^v, \boldsymbol{\delta}^v)$ as the maximal admissible carried load $\rho^v$ given a fixed relative load distribution $\tilde{\boldsymbol{\rho}}^v$, BTD requirement $d_v$, and mean reciprocal capacities $\boldsymbol{\delta}^v$. As might be expected, if the relative load distribution $\tilde{\boldsymbol{\rho}}^v$ is more balanced (normalized by the mean base station capacity), i.e., $\|\tilde{\boldsymbol{\rho}}^v\|_{\boldsymbol{\Delta}^v}^2$ is smaller, or if the BTD constraint is relaxed, i.e., $d_v$ is higher, or the base station capacities scale up, i.e., $\boldsymbol{\delta}^v$ is smaller, the slice can carry a higher overall load $\rho^v$.

Next, let us consider SCPF based sharing amongst a set of slices $\mathcal{V}$ each with its own BTD requirements. It follows from Eq. (5) that to meet such requirements on each slice the following should hold: for all $v\in\mathcal{V}$

$$s^v \geq \frac{1+\rho^v}{l(d_v,\tilde{\boldsymbol{\rho}}^v,\boldsymbol{\delta}^v) - \rho^v} \sum_{u\neq v} s^u \frac{\langle \tilde{\boldsymbol{\rho}}^v, \tilde{\boldsymbol{\rho}}^u\rangle_{\boldsymbol{\Delta}^v}}{\|\tilde{\boldsymbol{\rho}}^v\|_{\boldsymbol{\Delta}^v}^2}. \quad (15)$$

This can be written as:

$$\sum_{v\in\mathcal{V}} s^v \boldsymbol{h}^v \succeq \boldsymbol{0}, \quad (16)$$

where we refer to $\boldsymbol{h}^v = (h_u^v : u \in \mathcal{V})$ as $v$'s *share coupling vector*, given by

$$h_u^v = \begin{cases} 1 & v = u \\ -\frac{1+\rho^u}{l(d_u,\tilde{\boldsymbol{\rho}}^u,\boldsymbol{\delta}^u)-\rho^u}\frac{\langle \tilde{\boldsymbol{\rho}}^u, \tilde{\boldsymbol{\rho}}^v\rangle_{\boldsymbol{\Delta}^u}}{\|\tilde{\boldsymbol{\rho}}^u\|_{\boldsymbol{\Delta}^u}^2} & v \neq u, \end{cases}$$

We can interpret $h_v^v = 1$ as the benefit to slice $v$ of allocating unit share to itself. When $v \neq u$, $h_u^v$ depends on two factors. The first $\frac{1+\rho^u}{l(d_u,\tilde{\boldsymbol{\rho}}^u,\boldsymbol{\delta}^u)-\rho^u}$ captures the sensitivity of slice $u$ to the 'share weighted congestion' from other slices. If $\rho^u$ is close to its limit $l(d_u, \tilde{\boldsymbol{\rho}}^u, \boldsymbol{\delta}^u)$, its sensitivity is naturally very high. The second term, $\frac{\langle \tilde{\boldsymbol{\rho}}^u, \tilde{\boldsymbol{\rho}}^v\rangle_{\boldsymbol{\Delta}^u}}{\|\tilde{\boldsymbol{\rho}}^u\|_{\boldsymbol{\Delta}^u}^2}$ captures the impact of slice $v$'s load distribution on slice $u$. Note that if two slices load distributions are orthogonal, they do not affect each other.

The following result summarizes the above analysis.

**Theorem 2.** *There exists a share allocation such that slice loads and BTD constraints $((\rho^v, \tilde{\boldsymbol{\rho}}^v, d_v) : v \in \mathcal{V})$ are admissible under SCPF sharing if and only if there exists an $\boldsymbol{s} = (s^v : v \in \mathcal{V})$ such that $\|\boldsymbol{s}\|_1 = 1$, $\boldsymbol{s} \succeq \boldsymbol{0}$ and*

$$\sum_{v\in\mathcal{V}} s^v \boldsymbol{h}^v \succeq 0.$$

Admissibility can then be verified by solving the following maxmin problem:

$$\max_{\boldsymbol{s}\succeq\boldsymbol{0}}\{\ \min_i \sum_{v\in\mathcal{V}} s^v h_i^v : \|\boldsymbol{s}\|_1 = 1\ \}. \quad (17)$$

If the optimal objective function is positive, the traffic pattern is admissible. Moreover, if there are multiple feasible share allocations, then the optimizer is a 'robust' choice in that it maximizes the minimum share given to any slice, giving slices margins to tolerate perturbations in the slice loads satisfying Eq. (16).

If a set of network slice loads and BTD constraints are not admissible, admission control will need to be applied. We discuss this in the next section.

## V. Admission Control and Traffic Shaping Games

A natural approach to managing performance in overloaded systems is to perform admission control. In the context of slices supporting mobile services where spatial loads may vary substantially, this may be unavoidable. Below we consider admission control policies that adapt to changes in load. Specifically, an *admission control policy* for slice $v$ is parameterized by $\boldsymbol{a}^v \triangleq (a_b^v : b \in \mathcal{B}) \in [0,1]^B$ where $a_b^v$ is the probability a new customer at base station $b$ is admitted. Such decisions are assumed to be made independently thus admitted customers for slice $v$ at base station $b$ still follow a Poisson Process with rate $\gamma_b^v a_b^v$. Based on the flow conservation equation Eq. (1) one can obtain the carried load $\boldsymbol{\rho}^v$ induced by admission control policy $\boldsymbol{a}^v$ via

$$\boldsymbol{\rho}^v = (\boldsymbol{M}^v)^{-1} \boldsymbol{a}^v = \text{diag}(\boldsymbol{\mu}^v)(\boldsymbol{I} - (\boldsymbol{Q}^v)^T)^{-1} \text{diag}(\boldsymbol{\gamma}^v) \boldsymbol{a}_v$$

where $\boldsymbol{M}^v \triangleq \text{diag}(\boldsymbol{\gamma}^v)^{-1}(\boldsymbol{I} - (\boldsymbol{Q}^v)^T)\text{diag}(\boldsymbol{\mu}^v)^{-1}$ is invertible because $\boldsymbol{I} - (\boldsymbol{Q}^v)^T$ is irreducibly diagonally dominant.[1] By contrast with Section II-A, note that $\boldsymbol{\rho}^v$ now represents the load after admission control, which may have a reduced overall load and possibly changed relative loads across base stations–i.e., *shape* the traffic on the slice. We also let $\tilde{\boldsymbol{g}}$ be the overall share weighted relative loads after admission control, see Section III-A. Note that we have assumed only exogenous arrivals can be blocked, thus once a customer is admitted it will not be dropped–the intent is to manage performance to maintain *service continuity*.

Below we consider a setting where slices *unilaterally* optimize their admission control policies in response to network congestion, rather than a single joint global optimization. The intent is to allow slices (which may correspond to competing virtual operators/services) to optimize their own performance, and/or enable decentralization in settings with SCPF based sharing.

For simplicity we assume that assumption 1 holds true throughout this section, and define the capacity normalized mean BTD requirement $\tilde{d}_v \triangleq \frac{d_v}{\delta_v}$. Suppose each slice $v$ optimizes its admission control policy so as to maximize its overall carried load $\rho^v$, i.e., the average number of active users on the network, subject to a normalized mean BTD constraint $\tilde{d}_v$. Under Assumption 1 the optimal policy for slice $v$ is the solution to the following optimization problem:

$$\max_{\tilde{\boldsymbol{\rho}}^v, \rho^v} \rho^v \quad (18)$$

$$\text{s.t.} \quad \boldsymbol{a}^v = \rho^v \boldsymbol{M}^v \tilde{\boldsymbol{\rho}}^v, \quad \boldsymbol{a}^v \in [0,1]^B, \quad \langle \mathbf{1}, \tilde{\boldsymbol{\rho}}^v \rangle = 1 \quad (19)$$

$$\frac{(\rho^v + 1)}{s^v} \langle \tilde{\boldsymbol{g}}, \tilde{\boldsymbol{\rho}}^v \rangle - \left(1 - (\rho^v + 1)e^{-\rho^v}\right) \|\tilde{\boldsymbol{\rho}}^v\|_2^2$$
$$\leq \tilde{d}_v - 1 \quad (20)$$

Note that Eq. (19) establishes a one-to-one mapping between $(\tilde{\boldsymbol{\rho}}^v, \rho^v)$ and $\boldsymbol{a}^v$. We will use $\tilde{\boldsymbol{\rho}}^v$ and $\rho^v$ to parameterize admission control decisions for slice $v$. The BTD constraint in

[1] If $\boldsymbol{\gamma}^v$ is not strictly positive one can reduce the dimensionality.

Eq. (20) follows from Eq. (5). Also note that this admission control policy depends on both the overall share weighted loads on the network $\tilde{\boldsymbol{g}}$, the slice's load and its customer mobility patterns (i.e., $\boldsymbol{M}^v$). Unfortunately, for general loads $\rho^v$, this problem is not convex due to the BTD constraint Eq. (20); however, for high overall per slice loads it is easily approximable by a convex function.

Under Assumption 2 we have that $1 + \rho^v \approx \rho^v$ and the left hand side of Eq. (20) becomes:

$$\frac{(\rho^v + 1)}{s^v}\langle \tilde{\boldsymbol{g}}, \tilde{\boldsymbol{\rho}}^v \rangle - \|\tilde{\boldsymbol{\rho}}^v\|_2^2 \approx \frac{\rho^v}{s^v} \langle \tilde{\boldsymbol{g}}, \tilde{\boldsymbol{\rho}}^v \rangle = (s^v x_v)^{-1} \langle \tilde{\boldsymbol{g}}, \tilde{\boldsymbol{\rho}}^v \rangle \quad (21)$$

where we have defined $x_v \triangleq (\rho^v)^{-1}$. Further defining $\tilde{\boldsymbol{\rho}}^{-v} \triangleq (\tilde{\boldsymbol{\rho}}^{v'} : v' \in \mathcal{V} \setminus \{v\})$, Eq. (20) can be replaced by:

$$f_v(\tilde{\boldsymbol{\rho}}^v; \tilde{\boldsymbol{\rho}}^{-v}) \triangleq \langle \tilde{\boldsymbol{g}}, \tilde{\boldsymbol{\rho}}^v \rangle \leq s^v(\tilde{d}_v - 1) x_v. \quad (22)$$

Thus, by defining $\boldsymbol{y}^v \triangleq (\tilde{\boldsymbol{\rho}}^v, x^v)$, which is equivalent to $(\tilde{\boldsymbol{\rho}}^v, \rho^v)$, together with $\boldsymbol{y}^{-v} \triangleq (\boldsymbol{y}^{v'} : v' \in \mathcal{V} \setminus \{v\})$, each slice can unilaterally optimize its admission control policy by solving the following problem:

**Admission control for slice $v$ under SCPF ($\text{AC}_v$):** Given other slices' admission decisions $\boldsymbol{y}^{-v}$, slice $v$ determines its admission control policy $\boldsymbol{y}^v = (\tilde{\boldsymbol{\rho}}^v, x^v)$ by solving

$$\min_{\boldsymbol{y}^v} \{ x_v \mid \boldsymbol{y}^v \in Y^v(\boldsymbol{y}^{-v}) \} \quad (23)$$

where $Y^v(\boldsymbol{y}^{-v})$ denotes slice $v$'s feasible policies and is given by

$$Y^v(\boldsymbol{y}^{-v}) \triangleq \{ \boldsymbol{y}^v \mid \langle \mathbf{1}, \tilde{\boldsymbol{\rho}}^v \rangle = 1, \ \mathbf{0} \preceq \boldsymbol{M}^v \tilde{\boldsymbol{\rho}}^v \preceq x_v \mathbf{1},$$
$$f_v(\tilde{\boldsymbol{\rho}}^v; \tilde{\boldsymbol{\rho}}^{-v}) \leq s^v(\tilde{d}_v - 1) x_v \}. \quad (24)$$

Note that $\text{AC}_v$ is coupled to the decisions of other slices through the feasible set $Y_v(\boldsymbol{y}^{-v})$. Thus, one cannot independently solve each slice's admission control problem to obtain an efficient solution. Furthermore, devising a global optimization for all slices brings both complexity and nonconvexity from the BTD constraints. A natural approach requiring minimal communication and cooperation overhead is to consider a game setup where network slices are players, each seeking to maximize their carried loads (and the corresponding revenue) subject to BTD constraints.

We formally define the traffic shaping game for a set of network slices $\mathcal{V}$ as follows. We let $\boldsymbol{y} \triangleq (\boldsymbol{y}_v : v \in \mathcal{V})$ denote the simultaneous strategies of all slices (given by the respective admission control policies). As in $\text{AC}_v$, each slice $v$ picks a feasible strategy, i.e., $\boldsymbol{y}^v \in Y^v(\boldsymbol{y}^{-v})$ to minimize its objective function $\theta_v(\boldsymbol{y}^v, \boldsymbol{y}^{-v}) \triangleq x_v$. Note in the sequel we will modify $\theta_v(\cdot, \cdot)$ to ensure the games convergence. A Nash equilibrium is a simultaneous strategy $\boldsymbol{y}^*$ such that no slice can unilaterally improve its carried load, i.e., for all $v \in \mathcal{V}$

$$\theta_v(\boldsymbol{y}^{v,*}, \boldsymbol{y}^{-v,*}) \leq \theta_v(\boldsymbol{y}^v, \boldsymbol{y}^{-v,*}), \quad \forall \boldsymbol{y}^v \in Y^v(\boldsymbol{y}^{-v,*}).$$

The following result follows from Theorem 3.1 in [13].

**Theorem 3.** *The traffic shaping game defined above has a Nash equilibrium.*

Note that at the Nash equilibrium, no slice can unilaterally improve its performance. Therefore, finding the Nash equilibrium is also a way to achieve fairness under our sharing scheme. In the next subsection, we will design an algorithm to achieve such allocation.

*A. Algorithm*

In our setting finding the Nash equilibrium is not a simple matter. The difficulty arises from the fact that slices' strategy spaces depend on other's choices, so oscillation is possible. In the literature such settings are specifically referred to as Generalized Nash Equilibrium Problem (GNEP), see, e.g., [14] and [29]. However, the algorithm proposed in [14] assumed an algorithm capable of solving a penalized unconstrained Nash Equilibrium Problem, which satisfies a set of conditions, and that in [29] relies on the convexity of the joint strategy space. Thus none of them can be directly applied in our setting. Below we propose an algorithm involving slices and a central entity which is guaranteed to converge to the equilibrium.

We summarize the main ideas as follows. To decouple dependencies among strategy spaces, we shall move slice $v$'s BTD constraint into its objective function as a penalty term with an associated multiplier $\lambda_v$. Let $\boldsymbol{\lambda} \triangleq (\lambda_v : v \in \mathcal{V})$. By adjusting the value of $\boldsymbol{\lambda}$ according to $\boldsymbol{y}$ at each iteration, one can determine a setting such that, at the induced Nash Equilibrium, all slices meet their BTD constraints, and the equilibrium is identical to that of the traffic shaping game. In addition, in order to prevent overshooting, at each iteration each slice's objective function is regularized by the distance to the previous reciprocal carried load $x_v$.

Specifically, the admission control strategy of slice $v$ in response to other slices is now given as the solution to the following optimization problem:

$$L_\epsilon^v(\boldsymbol{y}; \lambda^v) = \operatorname*{argmin}_{(\boldsymbol{y}^v)' \in \bar{Y}^v} \theta_v((\boldsymbol{y}^v)', \boldsymbol{y}^{-v}; \lambda_v) + \frac{\epsilon}{2}(x_v - x_v')^2, \quad (25)$$

where we define a BTD penalty function for slice $v$ as

$$h_v(\boldsymbol{y}) \triangleq f_v(\tilde{\boldsymbol{\rho}}^v; \tilde{\boldsymbol{\rho}}^{-v}) - s^v(\tilde{d}_v - 1)x_v.$$

and the objective function for slice $v$ is now (different from what is previously defined): $\theta_v(\boldsymbol{y}^v, \boldsymbol{y}^{-v}; \lambda_v) \triangleq e^{x_v} + \lambda_v[h_v(\boldsymbol{y})]_+$, with $[x]_+ \triangleq \max(0, x)$. The last term in Eq. (25) serves as a regularization term. The strategy space is now $\bar{Y}^v \triangleq \{\boldsymbol{y}_v | \langle \boldsymbol{1}, \tilde{\boldsymbol{\rho}}^v \rangle = 1, \boldsymbol{0} \preceq \boldsymbol{M}^v \tilde{\boldsymbol{\rho}}^v \preceq x_v \boldsymbol{1}\}$ and $x_v$ is substituted by $e^{x_v}$ to ensure strong convexity, which is required for convergence (note that due to the monotonicity, $e^{x_v}$ and $x_v$ should result in the same optimizer).

We propose to use the inexact line search update introduced in [29]. In order to make sure the iteration is proceeding towards the equilibrium, we use

$$\Omega_\epsilon(\boldsymbol{y}; \boldsymbol{\lambda}) \triangleq \sum_{v \in \mathcal{V}} \theta_v(\boldsymbol{y}^v, \boldsymbol{y}^{-v}; \lambda_v) - \theta_v(L_\epsilon^v(\boldsymbol{y}; \boldsymbol{\lambda}), \boldsymbol{y}^{-v}; \lambda_v)$$
$$- \frac{\epsilon}{2}(x_v - x_v')^2 \geq 0$$

as a metric, observing that the equilibrium is given by $\boldsymbol{y}^*$ if and only if $\Omega_\epsilon(\boldsymbol{y}^*; \boldsymbol{\lambda}) = 0$. Therefore we seek to decrease $\Omega_\epsilon(\boldsymbol{y}; \boldsymbol{\lambda})$ by a sufficient amount at each iteration. The task executed by each slice $v$ is given in Algorithm 1, while the central entity, which is responsible for collecting and delivering information and updating $\boldsymbol{\lambda}$, executes Algorithm 2. This then follows the algorithm proposed in [14].

---

**Algorithm 1** Algorithm of Slice $v$

1: Set $k \leftarrow 0$ and collect $\epsilon$ from central entity.
2: Receive $\lambda_v(k)$ and $\boldsymbol{y}(k)$ from central entity.
3: Compute $L_\epsilon^v(\boldsymbol{y}(k); \boldsymbol{\lambda})$ and transmit it back to the central entity. Set $k \leftarrow k + 1$. Go to step 2

---

**Algorithm 2** Penalized Update in Central Entity

1: Choose a starting point $\boldsymbol{y}(0)$, $\boldsymbol{\lambda}(0) \succeq \boldsymbol{0}$, $\eta_v \in (0, 1)$, for $v \in \mathcal{V}$, $\beta, \sigma \in (0, 1)$, $\epsilon > 0$ but small enough (see following theorem for convergence) and set $k \leftarrow 0$.
2: If a termination criterion is met then STOP. Otherwise, communicate $\boldsymbol{y}(k)$ together with $\lambda_v(k)$ to all slices.
3: All slices compute $L_\epsilon^v(\boldsymbol{y}(k); \boldsymbol{\lambda})$ and feedback to central entity.
4: Compute $t(k) = \max\{\beta^l | l = 0, 1, 2, \dots\}$ such that if we assume $\xi(k) = (L_\epsilon^v(\boldsymbol{y}(k); \boldsymbol{\lambda}(k)) : v \in \mathcal{V}) - \boldsymbol{y}(k)$:

$$\Omega_\epsilon(\boldsymbol{y}(k) + t(k)\xi(k)) \leq \Omega_\epsilon(\boldsymbol{y}(k)) - \sigma(t(k))^2 \|\xi(k)\|. \quad (26)$$

Then set $\boldsymbol{y}(k+1) = \boldsymbol{y}(k) + t(k)\xi(k)$.
5: Set $I(k) = \{v | h_v(\boldsymbol{y}(k)) > 0\}$. For every $v \in I(k)$, if

$$e^{x_v(k)} > \eta_v(\lambda_v \|\nabla_{\boldsymbol{y}^v} h_v \boldsymbol{y}(k)\|), \quad (27)$$

then $\lambda_v(k+1) \leftarrow 2\lambda_v(k)$. Set $k \leftarrow k+1$. Broadcast $\boldsymbol{y}(k)$ and $\boldsymbol{\lambda}(k)$ to slices and go to step 2.

---

**Theorem 4.** *Let $\{\boldsymbol{y}(k)\}$ be the sequence of admission control decisions generated by Algorithm 1 and Algorithm 2, then every limit point of this sequence is a Nash equilibrium of the traffic shaping game induced by $AC_v$.*

*Proof.* First we need to verify the Assumption 5.1 in [29] to guarantee that for a given $\boldsymbol{\lambda}$, step 4 in Algorithm 2 converges to a Nash equilibrium. The non-constant part of $\Psi_\epsilon(\boldsymbol{y}, \boldsymbol{y}'; \boldsymbol{\lambda})$ (defined in [29]) when $\boldsymbol{y}'$ is fixed is: $\sum_v e^{x_v} + \lambda_v[h_v(\boldsymbol{y})]_+ - \frac{\epsilon}{2}\|x_v - x_v'\|^2$. If $\epsilon$ is small enough, the concavity of the last term will be canceled out by $e^{x_v}$. Then the non-constant part is always convex in $\boldsymbol{y}$. Hence, the Assumption 5.1 holds true together with the propositions 2.1(a) - (d) in [29]. Therefore, the proposed algorithm generates Nash equilibrium of the game.

One can easily verify that the EMFCQ condition given by Definition 2.7 in [14] is satisfied. Thus for all $v$, $\lambda_v$ gets updated a finite number of times. According to Theorem 2.5 in [14], the claim is true. □

*B. Characterization of Traffic Shaping Equilibrium*

Next we study the characteristics of the resulting traffic shaping Nash equilibrium. To make this tractable we consider networks which are saturated and subsequently (in Section VI) provide simulations to evaluate other settings.

**Assumption 3.** (Saturated Regime) *Suppose the system is such that for each network slice, the optimal admission control for both SCPF and SS [2] in response to other slices' loads is such that for all $v \in \mathcal{V}$, $\boldsymbol{a}^v \prec \boldsymbol{1}$.*

Assumption 3 depends on many factors including the BTD constraints, the mobility patterns, and network slices' shares, but it is generally true when the exogenous traffic of all slices at all base stations $\gamma_b^v$ is high. When this is the case we have the following result:

**Theorem 5.** *Under Assumptions 1, 2 and 3, the relative load distributions at the Nash equilibrium of the traffic shaping game $\tilde{\boldsymbol{\rho}}^* \triangleq (\tilde{\boldsymbol{\rho}}^{v,*} : v \in \mathcal{V})$ are the unique solution to:*

$$\min_{(\tilde{\boldsymbol{\rho}}^v \in \Gamma^v : v \in \mathcal{V})} \| \sum_v s^v \tilde{\boldsymbol{\rho}}^v \|_2^2 + \sum_v (s^v)^2 \|\tilde{\boldsymbol{\rho}}^v\|_2^2, \quad (28)$$

*where $\Gamma^v \triangleq \{ \tilde{\boldsymbol{\rho}}^v \mid \langle \boldsymbol{1}, \tilde{\boldsymbol{\rho}}^v \rangle = 1, \boldsymbol{M}^v \tilde{\boldsymbol{\rho}}^v \succeq \boldsymbol{0} \}$, and the associated carried load for slice $v$ is $\rho^{v,*} = \frac{s^v(\tilde{d}_v - 1)}{\langle \tilde{\boldsymbol{g}}^*, \tilde{\boldsymbol{\rho}}^{v,*} \rangle}$, where $\tilde{\boldsymbol{g}}^*$ corresponds to the overall share weighted relative loads distributions at the equilibrium.*

See appendix for detailed proof.

The first term in the objective function in Eq. (28) rewards balancing the overall share weighted relative loads on network. The second term rewards a slice for balancing its own relative loads. The Nash equilibrium in the saturated regime is thus a compromise between those two objectives while constrained by the network slices mobility patterns and feasible admission control policies. Note that as long as $\tilde{\rho}_b^v > 0, \forall v \in \mathcal{V}, b \in \mathcal{B}$, GPS and SS are approximately the same under heavy load. Therefore, we use SS as the benchmark to characterize the carried load at the Nash equilibrium under SCPF.

**Admission control for slice $v$ under SS (ACSS$_v$):** Under SS slice $v$ can determine its optimal admission control $\boldsymbol{y}^v$ by solving:

$$\max_{\tilde{\boldsymbol{\rho}}^v, \rho^v} \quad \rho^v$$
$$\text{s.t.} \quad \boldsymbol{a}^v = \rho^v \boldsymbol{M}^v \tilde{\boldsymbol{\rho}}^v, \quad \boldsymbol{a}^v \in [0,1]^B$$
$$\langle \boldsymbol{1}, \tilde{\boldsymbol{\rho}}^v \rangle = 1 \text{ and } \rho^v \|\tilde{\boldsymbol{\rho}}^v\|_2^2 \leq (s^v \tilde{d}_v - 1).$$

Note slices' admission control decisions are clearly decoupled under SS, but paralleling Theorem 5 we have following result.

**Theorem 6.** *Under Assumptions 1 and 3, the optimal admission control policy under SS are decoupled. The optimal choice for slice $v$, $\tilde{\boldsymbol{\rho}}^{v,SS,*}$, is the unique solution to:*

$$\min_{\tilde{\boldsymbol{\rho}}^v \in \Gamma^v} \|\tilde{\boldsymbol{\rho}}^v\|_2^2, \quad (29)$$

*and the associated carried load is given by $\rho^{v,SS,*} = \frac{s^v \tilde{d}_v - 1}{\|\tilde{\boldsymbol{\rho}}^{v,SS,*}\|_2^2}$.*

See appendix for detailed proof.

[2]Admission control under SS is defined in the sequel.

By comparing Eq. (28) and Eq. (29), one can see that under SS, slices simply seek to balance their own relative loads on the network. By taking the ratio between $\rho_v^*$ and $\rho_v^{SS,*}$ given in Theorem 5 and 6, one can show that under Assumptions 1, 2, and 3 the gain in carried load for slice $v$ is given by

$$G_v^{\text{load}} \triangleq \frac{\rho^{v,*}}{\rho^{v,SS,*}} = \frac{\|\tilde{\boldsymbol{\rho}}^{v,SS,*}\|_2^2}{\langle \tilde{\boldsymbol{g}}^*, \tilde{\boldsymbol{\rho}}^{v,*} \rangle} \times \frac{s^v(\tilde{d}_v - 1)}{s^v \tilde{d}_v - 1}. \quad (30)$$

The first factor captures a traffic shaping dependent gain for slice $v$. The second factor is a result of statistical multiplexing gains. A simple special case is highlighted in the following corollary.

**Corollary 4.** *Under Assumptions 1, 2 and 3, if user mobility patterns are such that $\frac{1}{B}\boldsymbol{1} \in \Gamma^v$ for all $v \in \mathcal{V}$, the gain in the total carried load under the SCPF traffic shaping Nash equilibrium vs. optimal admission control for SS is given by:*

$$G_v^{\text{load}} = \frac{s^v \tilde{d}_v - s^v}{s^v \tilde{d}_v - 1} \geq 1, \quad \forall v \in \mathcal{V}. \quad (31)$$

See appendix for detailed proof.

Note that in order for a BTD constraint to be feasible under SS, one must require $s^v \tilde{d}_v > 1$. It can be seen that the gain exhibited in Corollary 4 can be very high when $s^v \downarrow 1/\tilde{d}_v$. Furthermore, if $s^v \uparrow 1$ we have that $G_v^{\text{load}} \downarrow 1$, i.e., no actual gain. This result implies that slices with small shares or tight BTD constraints will benefit most from sharing, coinciding with our observations in Corollary 1.

## VI. PERFORMANCE EVALUATION RESULTS

In this section, we validate theoretical results in previous sections, and provide quantitative characterizations via numerical experiments. We simulated a wireless network shared by multiple slices supporting mobile customers following the IMT-Advanced evaluation guidelines [18]. The system consists of 19 base stations in a hexagonal cell layout with an inter site distance of 200 meters and 3 sector antennas, mimicking a dense 'small cell' deployment. Thus, in this system, $\mathcal{B}$ corresponds to 57 sectors. Users associate to the sector offering the strongest SINR, where the downlink SINR is modeled as in [31]:

$$\text{SINR}_{ub} = \frac{P_b G_{ub}}{\sum_{k \in \mathcal{B}, k \neq b} P_k G_{uk} + \sigma^2},$$

where, following [18], the noise $\sigma^2 = -104$dB, the transmit power $P_b = 41$dB and the channel gain between user $u$ and BS sector $b$, denoted by $G_{ub}$, accounts for path loss, shadowing, fast fading and antenna gain. Letting $d_{ub}$ denote the current distance in meters from the user $u$ to sector $b$, the path loss is defined as $36.7 \log_{10}(d_{ub}) + 22.7 + 26 \log_{10}(f_c)$dB, for a carrier frequency $f_c = 2.5$GHz. The antenna gain is set to 17 dBi, shadowing is updated every second and modeled by a log-normal distribution with standard deviation of 8dB, as in [31]; and fast fading follows a Rayleigh distribution depending on the mobile's speed and the angle of incidence. The downlink rate $c_u$ currently achievable to user $u$ is based on discrete set modulation and coding schemes (MCS) and associated SINR thresholds given in [3]. This MCS value is selected based on the averaged $\overline{\text{SINR}}_{ub}$, where channel fast fading is averaged over a second.

| Scenario | Slices | Spatial loads | $\|\tilde{\boldsymbol{\rho}}^v\|_2$ | $\|\tilde{\boldsymbol{g}}\|_2$ | $\theta(\tilde{\boldsymbol{g}}, \tilde{\boldsymbol{\rho}}^v)$ | $G_v^{SS,H}$ |
|---|---|---|---|---|---|---|
| 1 | Homogeneous | uniform. | 0.27 | 0.27 | 7.09 | 1.01 |
| 2 | Homogeneous | non-uniform | 0.32 | 0.32 | 6.18 | 1.01 |
| 3 | Heterogeneous | orthogonal | 0.36 | 0.26 | 41.78 | 1.83 |
| 4 | Mixed Slices | 1&2 non-uniform | 0.36 | 0.23 | 25.52 | 1.70 |
|   |   | 3&4 uniform | 0.19 | 0.23 | 48.00 | 1.24 |

TABLE II: Measured normalized slice and network traffic norms and angles for highest load case of each scenario.

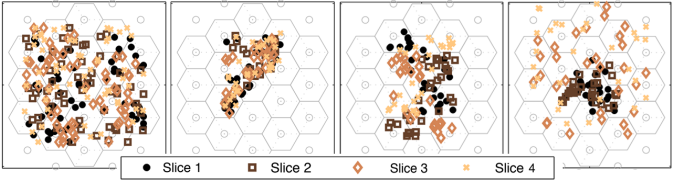

Fig. 1: Snapshot of users positions per slice and scenario exhibiting the different characteristics of traffic spatial loads. Left to right: Scenarios 1 to 4.

We model slices' with different spatial loads by modeling different customer mobility patterns. Roughly uniform spatial loads are obtained by simulating the Random Waypoint model [17], while non-uniform loads obtained by simulating the SLAW model [21]. Instead of the open network assumed in the theoretical analysis, in the simulations we use a closed network where the total number of users on each slice keeps fixed. Moreover, the simulated mobility models would not induce Markovian motion amongst base stations assumed in our analysis, yet the analytical results are robust to these assumptions.

### A. Statistical Multiplexing and BTD Gains

We evaluated the BTD gains of SCPF vs. both SS and GPS for four simulation scenarios, each including 4 slices, each with equal shares but different spatial load patterns. For each scenario, we provide results for simulated BTD gains, and results from our theoretical analysis (Corollary 1 and Corollary 2) based on the empirically obtained spatial traffic loads. More detailed information regarding simulated scenarios and resulting empirical spatial traffic loads for high load regime are displayed in Table II and a snapshot of locations for the 4 slices' users in a network with a load of 4 users per sector is displayed in Figure 1.

The results given in Figure 2 show the BTD gains over SS for each scenario as the overall network load increases. In Scenario 3, the aggregate network traffic is 'smoother' than the individual slice's traffic, and the gains are indeed higher. This is also the case for Slice 1 and 2 in Scenario 4, since these slices loads are more 'imbalanced' than the other two slices, they experience higher gains. In Scenario 2, where slices non-homogenous spatial loads are 'aligned', aggregation does not lead to smoothing and the gains are least.

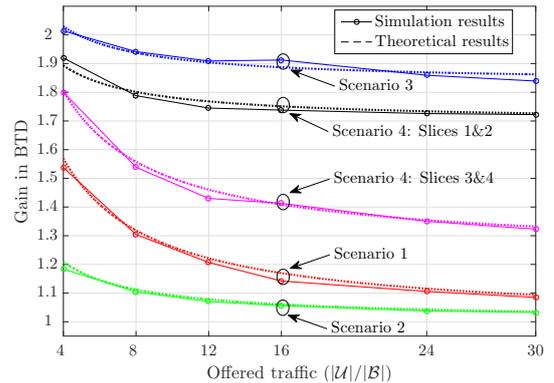

Fig. 2: BTD gain over SS for our 4 different scenarios.

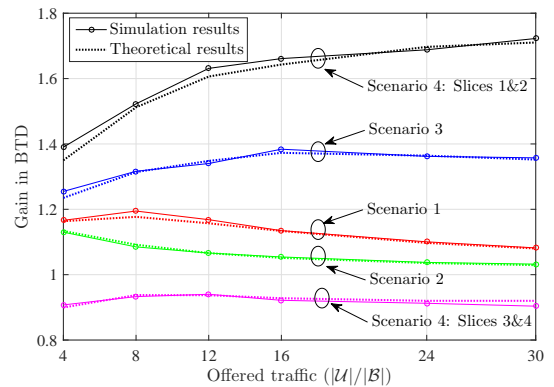

Fig. 3: BTD gain over GPS for our 4 different scenarios.

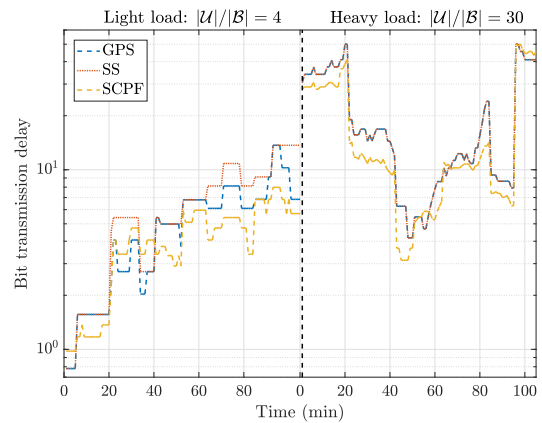

Fig. 4: BTD vs. time for a randomly picked user under Scenario 3

Similarly, the results given in Figure 3 show the BTD gains over GPS for each scenario as the overall network load increases. As can be seen, the gain is not necessarily monotonic in the load. In Scenario 4, the Slice 1 and 2 have significant gains because their loads are more imbalanced, while Slice 3 and 4 see negative gains. However, the overall gain defined in Eq. (13) is still positive, ranging from 1.26 to 1.5 for varying overall load.

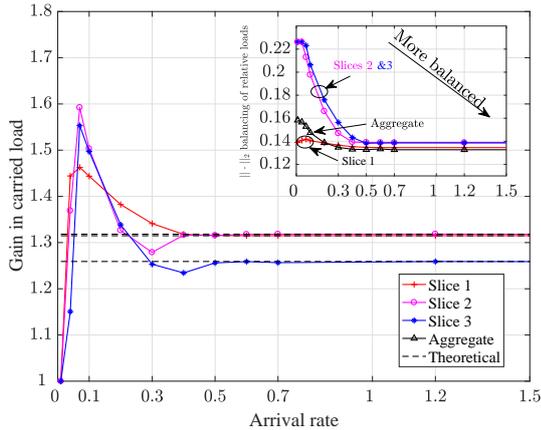

Fig. 5: Gain in carried load for various arrival rates. Subfigure: Balancing in relative load.

As can be seen in Figures 2 and 3 the simulated and theoretical gains (dashed lines) of Corollary 1 are an excellent match. The theoretical model has been calibrated to the mean reciprocal capacities seen by slice customers (i.e., $\delta_b^v$'s) and the measured induced loads resulting from the slice mobility patterns.

In addition to performance averaged over time, to illustrate the dynamic of the BTD perceived by a typical user, we plot the BTD vs. time for a randomly picked user on Slice 1 in Scenario 1, as shown in Fig. 4, where the left/right part is under light/heavy load regime, respectively. Note that under heavy load, GPS and SS are approximately the same. SCPF outperforms for most of the time. Under light load, the mean BTD under SCPF is 4.2044, while that under SS (GPS) is 6.6157 (5.2862), respectively. The standard deviation of BTD under SS (GPS) is 3.9011 (3.0449), and SCPF reduces it to 1.93. Similar phenomenon is observed under heavy load, when both SS and GPS provide mean BTD of 19.65 and associated standard deviation of 13.79, SCPF reduce them to 16.79 and 13.45, respectively. Therefore, SCPF can effectively improve the perceived BTD and also 'smooth' the user perceived QoS.

### B. Traffic Shaping Equilibrium and Carried Load Gains

In order to study the equilibria reached by the traffic shaping game, we measured the underlying user mobility patterns in Section VI-A, and modeled it via a random routing matrix. We further assumed uniform intensity of arrivals rates at all base stations and uniform exit probabilities of 0.1. The mean holding time at each base station was again calibrated with the simulations in Section VI-A. We considered a traffic shaping game for a network shared by 3 slices, where Slice 1 has uniform spatial loads and Slice 2 and 3 have different non-uniform spatial loads. All slices have equal shares and their capacity normalized BTD requirements are set to $\tilde{d}_1 = 10, \tilde{d}_2 = 12, \tilde{d}_3 = 15$ respectively. The Nash equilibrium was solved via the algorithm included in Section V-A. The convergence is reached within 3 rounds of iterations under the parameters $\eta_v = \beta = 0.5, \forall v \in \mathcal{V}, \sigma = 0.1, \epsilon = 0.01$. The results shown in Figure 5 exhibit dashed lines corresponding to the theoretical carried load gains in the saturated regime. As can be seen, these coincide with the Nash equilibria of the simulated traffic shaping games for high arrival rates. For lower arrival rates the gains can be much higher, e.g., almost a factor of 1.6, for slices with non-uniform mobility patterns. This was to be expected since for lower loads we expect higher statistical multiplexing gains from sharing, and thus relatively higher carried loads to be admitted. For very low loads, as expected, there are no gains since all traffic can be admitted and BTD constraints are met.

Also shown in Figure 5(subfigure) is the degree to which the relative loads of slices, and the weighted aggregate traffic on the network $\tilde{g}$ are balanced, as measured by $||\cdot||_2$, as the arrival rates on the network increase. As expected, based on Theorem 5, as arrivals increase relative loads of slices and the network become more balanced, showing the compromise the traffic shaping game is making, balancing slices relative loads and that of the overall network.

## VII. CONCLUSIONS

This paper has thoroughly explored a relatively simple and natural approach for resource sharing amongst network slices – SCPF – which corresponds to socially optimal allocations in a Fisher market. Our analysis of performance in settings where slices support stochastic loads provides explicit formulas for ($i$) the performance gains one can expect over SS and GPS, ($ii$) how to dimension slice shares to meet performance objectives, and ($iii$) how to go about performance management through admission control. If dynamic resource sharing amongst network slices is to be adopted, the ability to realize disciplined engineering and performance prediction will be the key. Our analysis of SCPF seems to meet these requirements and at the same time reveals some intriguing insights regarding the load interactions in such sharing models, in particular the impact of relative load distributions on statistical multiplexing, and the role of traffic shaping in optimizing admission control. Finally, we note that our approach to admission control in an SCPF shared system is novel in that each slice exploits knowledge of its customers' mobility patterns to optimize its carried load and assure service continuity.

## VIII. Appendix

### A. Proof of Theorem 1

*Proof.* Recall that Poisson arrivals see time averages, i.e., see the remaining users in the product-form stationary distribution, given in Section II-A. Thus the distribution as seen by a typical user on slice $v$ at base station $b$ is the same as the product-form distribution *plus an additional customer* on slice $v$ at base station $b$. Using this fact and SCPF resource allocations as given by Eq. (3), the mean BTD of a typical slice $v$ user at base station $b$ can be expressed as follows:

$$\mathbb{E}^v\left[\frac{1}{R_b^v}\right] = \mathbb{E}^v\left[\frac{1}{C_b^v}\right]\mathbb{E}\left[\frac{s^v\frac{N_b^v+1}{N^v+1} + \sum_{v'\neq v}\frac{s^{v'}N_b^{v'}}{N^{v'}}\mathbf{1}_{\{N^{v'}>0\}}}{\frac{s^v}{(N^v+1)}}\right]$$

$$= \delta_b^v \mathbb{E}\left[(N_b^v+1) + \frac{N^v+1}{s^v}\sum_{v'\neq v}\frac{s^{v'}N_b^{v'}}{N^{v'}}\mathbf{1}_{\{N^{v'}>0\}}\right]$$

$$= \delta_b^v(\rho_b^v + 1 + \frac{\rho^v+1}{s^v}\sum_{v'\neq v}s^{v'}(1-e^{\rho^{v'}})\tilde{\rho}_b^{v'})$$

$$= \delta_b^v(1 - \tilde{\rho}_b^v + \frac{(\rho^v+1)}{s^v}\tilde{g}_b' + (\rho^v+1)e^{-\rho^v}\tilde{\rho}_b^v),$$

where the second equality follows by noticing that (i) $N^v$ is independent of $N_b^{v'}$ and $N^{v'}$ and (ii) $E[\frac{N_b^{v'}}{N^{v'}}\mathbf{1}_{\{N^{v'}>0\}}] = P(N^{v'}>0)E[\frac{N_b^{v'}}{N^{v'}}|N^{v'}>0] = \frac{\rho_b^{v'}}{\rho^{v'}}P(N^{v'}>0)$. The latter result is given by the following lemma:

**Lemma 1.** *If $N_1, N_2, \ldots, N_n$ are independent Poisson random variables, such that $N_i \sim Poisson(\rho_i), i = 1, 2, \ldots, n$. Then for all $i$ we have that:*

$$\mathbb{E}\left[\frac{N_i}{\sum_{j=1}^n N_j}|\sum_{j=1}^n N_j > 0\right] = \frac{\rho_i}{\sum_{j=1}^n \rho_j}$$

*Proof.* Suppose all $\rho_i$'s are rational, and for some $\epsilon$ small enough, for all $i$, we have that $m_i = \frac{\rho_i}{\epsilon}$ is integer valued. Let $X_{i,j}, i = 1, \ldots, n, j = 1, \ldots, m_i$ be i.i.d. Poisson random variables with parameter $\epsilon$. Since Poisson random variables are infinitely divisible we have that $N_i \sim \sum_{j=1}^{m_i} X_{i,j}$. Then

$$\mathbb{E}\left[\frac{N_i}{\sum_{j=1}^n N_j}|\sum_{j=1}^n N_j > 0\right]$$

$$= \mathbb{E}\left[\frac{\sum_{j=1}^{m_i} X_{i,j}}{\sum_{i=1}^n \sum_{j=1}^{m_i} X_{i,j}}|\sum_{j=1}^n N_j > 0\right]$$

$$= m_i \cdot \mathbb{E}\left[\frac{X_{i,1}}{\sum_{i=1}^n \sum_{j=1}^{m_i} X_{i,j}}|\sum_{j=1}^n N_j > 0\right]$$

$$= \frac{m_i}{\sum_{j=1}^n m_j} = \frac{\rho_i}{\sum_{j=1}^n \rho_j},$$

where the second and third equalities follow from the symmetry among $X_{i,j}$. Since the conditional expectation will be a continuous function of the parameter vector $\rho$, the equality follows more generally for the case where $\rho$ is a real valued vector. □

The asymptotic form given in Eq. (5) follows by noting that when $\rho^v$ is large for all $v \in \mathcal{V}$, $\tilde{g}' \cong \tilde{g}$, and only the term scaling with $\rho^v$ matters.

Under static slicing we have that for a typical user on slice $v$ at base station $b$,

$$\mathbb{E}^v\left[\frac{1}{R_b^{v,SS}}\right] = \mathbb{E}^v\left[\frac{1}{C_b^v}\right]\mathbb{E}\left[\frac{N_b^v+1}{s^v}\right] = \delta_b^v \frac{\rho_b^v+1}{s^v}.$$

Similarly, under GPS we have that,

$$\begin{aligned}\mathbb{E}^v\left[\frac{1}{R^{v,GPS}}\right] &= \mathbb{E}^v\left[\frac{1}{C_b^v}\right]\mathbb{E}^v\left[\frac{N_b^v \sum_{v'\in\mathcal{V}} s^{v'}\mathbf{1}_{\{N_b^{v'}>0\}}}{s^v}\right] \\ &= \delta_b^v\left(\frac{\rho_b^v+1}{s^v}\right)(1-\bar{s}_b^v).\end{aligned}$$

The theorem follows by taking a weighted average across base stations – weighted by the fraction of customers at each base station, i.e., $\tilde{\rho}_b^v$. $\square$

*B. Proof of Corollary 1*

From Theorem 1 we have that for SCPF

$$\begin{aligned}\mathbb{E}^v\left[\frac{1}{R^v}\right] &= \langle\boldsymbol{\delta}^v,\tilde{\boldsymbol{\rho}}^v\rangle - \left(1-(\rho^v+1)e^{-\rho^v}\right)\|\tilde{\boldsymbol{\rho}}^v\|_{\boldsymbol{\Delta}^v}^2 \\ &\quad + \frac{\rho^v+1}{s^v}\langle\tilde{\boldsymbol{g}}',\tilde{\boldsymbol{\rho}}^v\rangle_{\boldsymbol{\Delta}^v},\end{aligned} \quad (32)$$

while for SS we have that

$$\mathbb{E}^v\left[\frac{1}{R^{v,SS}}\right] = \frac{1}{s^v}\left(\rho^v\|\tilde{\boldsymbol{\rho}}^v\|_{\boldsymbol{\Delta}^v}^2 + \langle\boldsymbol{\delta}^v,\tilde{\boldsymbol{\rho}}^v\rangle\right). \quad (33)$$

Taking the ratio of the overall mean BTDs we have Eq. (8) in Corollary 1.

Now setting $\rho^v = 0$, it is easy to see that

$$\begin{aligned}G_v^{SS,L} &= \frac{\langle\boldsymbol{\delta}^v,\tilde{\boldsymbol{\rho}}^v\rangle}{s^v\langle\boldsymbol{\delta}^v,\tilde{\boldsymbol{\rho}}^v\rangle + \langle\tilde{\boldsymbol{g}}',\tilde{\boldsymbol{\rho}}^v\rangle_{\boldsymbol{\Delta}^v}} \\ &= \frac{1}{s^v + \frac{\langle\tilde{\boldsymbol{g}}',\tilde{\boldsymbol{\rho}}^v\rangle_{\boldsymbol{\Delta}^v}}{\langle\boldsymbol{\delta}^v,\tilde{\boldsymbol{\rho}}^v\rangle}} \geq \frac{1}{s^v + 1 - s^v} = 1,\end{aligned}$$

Note that when $\rho^v \approx 0$, $\tilde{g}_b' \approx \sum_{v'\neq v} s^{v'}(1-e^{-\rho^{v'}})\tilde{\rho}_b^{v'}$. Therefore, the inequality follows from

$$\begin{aligned}\frac{\langle\tilde{\boldsymbol{g}}',\tilde{\boldsymbol{\rho}}^v\rangle_{\boldsymbol{\Delta}^v}}{\langle\boldsymbol{\delta}^v,\tilde{\boldsymbol{\rho}}^v\rangle} &= \frac{\sum_{b\in\mathcal{B}}\left(\sum_{v'\neq v}s^{v'}\left(1-e^{-\rho^{v'}}\right)\tilde{\rho}_b^{v'}\right)\delta_b^v\tilde{\rho}_b^v}{\sum_{b\in\mathcal{B}}\delta_b^v\tilde{\rho}_b^v} \\ &\leq \frac{\sum_{b\in\mathcal{B}}\left(\sum_{v'\neq v}s^{v'}\tilde{\rho}_b^{v'}\right)\delta_b^v\tilde{\rho}_b^v}{\sum_{b\in\mathcal{B}}\delta_b^v\tilde{\rho}_b^v} \\ &\leq \frac{\sum_{v'\neq v}s^{v'}\left(\sum_{b\in\mathcal{B}}\tilde{\rho}_b^v\delta_b^v\right)}{\sum_{b\in\mathcal{B}}\delta_b^v\tilde{\rho}_b^v} = 1-s^v.\end{aligned}$$

The last inequality follows from swapping the order of summation and $\tilde{\rho}_b^{v'} \leq 1, \forall b \in \mathcal{B}, v' \in \mathcal{V}$.

Let $\tilde{\boldsymbol{g}}_{-v}' = \sum_{v'\neq v} s^{v'}(1-e^{-\rho^{v'}})\tilde{\boldsymbol{\rho}}^{v'} + s^v\tilde{\boldsymbol{\rho}}^v$. Eq. (8) can be written as:

$$\begin{aligned}G_v^{SS} &= \frac{\|\tilde{\boldsymbol{\rho}}^v\|_{\boldsymbol{\Delta}^v}^2}{\langle\tilde{\boldsymbol{g}}_{-v}',\tilde{\boldsymbol{\rho}}^v\rangle_{\boldsymbol{\Delta}^v}} \\ &\quad + \langle\boldsymbol{\delta}^v,\tilde{\boldsymbol{\rho}}^v\rangle\frac{1-\left(\frac{\|\tilde{\boldsymbol{\rho}}^v\|_{\boldsymbol{\Delta}^v}^2}{\langle\boldsymbol{\delta}^v,\tilde{\boldsymbol{\rho}}^v\rangle}+(1-\frac{\|\tilde{\boldsymbol{\rho}}^v\|_{\boldsymbol{\Delta}^v}^2}{\langle\boldsymbol{\delta}^v,\tilde{\boldsymbol{\rho}}^v\rangle})\frac{s^v\|\tilde{\boldsymbol{\rho}}^v\|_{\boldsymbol{\Delta}^v}^2}{\langle\tilde{\boldsymbol{g}}_{-v}',\tilde{\boldsymbol{\rho}}^v\rangle_{\boldsymbol{\Delta}^v}}\right)}{(\rho^v+1)\langle\tilde{\boldsymbol{g}}_{-v}',\tilde{\boldsymbol{\rho}}^v\rangle_{\boldsymbol{\Delta}^v}+s^v(\langle\boldsymbol{\delta}^v,\tilde{\boldsymbol{\rho}}^v\rangle-\|\tilde{\boldsymbol{\rho}}^v\|_{\boldsymbol{\Delta}^v}^2)}.\end{aligned} \quad (34)$$

Note that $\langle\boldsymbol{\delta}^v,\tilde{\boldsymbol{\rho}}^v\rangle \geq \|\tilde{\boldsymbol{\rho}}^v\|_{\boldsymbol{\Delta}^v}^2$. Then because $\frac{s^v\|\tilde{\boldsymbol{\rho}}^v\|_{\boldsymbol{\Delta}^v}^2}{\langle\tilde{\boldsymbol{g}}_{-v}',\tilde{\boldsymbol{\rho}}^v\rangle_{\boldsymbol{\Delta}^v}} \leq 1$, the numerator of the second term is nonnegative. Therefore, $G_v^{SS}$ is decreasing in $\rho^v$. When $\rho^v \to \infty, \forall v \in \mathcal{V}$, the second term in Eq. (34) vanishes, and $\tilde{\boldsymbol{g}}_{-v}' \to \tilde{\boldsymbol{g}}$. Then $G_v^{SS,H}$ is given by

$$G_v^{SS,H} = \frac{\|\tilde{\boldsymbol{\rho}}^v\|_{\boldsymbol{\Delta}^v}^2}{\langle\tilde{\boldsymbol{g}},\tilde{\boldsymbol{\rho}}^v\rangle_{\boldsymbol{\Delta}^v}}. \quad (35)$$

*C. Proof of Corollary 2*

The BTD under SCPF is given in Eq. (32). Similarly for GPS we have that

$$\begin{aligned}\mathbb{E}^v\left[\frac{1}{R^{v,GPS}}\right] &= \\ \frac{1}{s^v}&\left[\rho^v(\|\tilde{\boldsymbol{\rho}}^v\|_{\boldsymbol{\Delta}^v}^2 - \|\tilde{\boldsymbol{\rho}}^v\|_{\boldsymbol{\Delta}^v\boldsymbol{S}^v}^2) + \langle\tilde{\boldsymbol{\rho}}^v,\mathbf{1}-\bar{\boldsymbol{s}}^v\rangle_{\boldsymbol{\Delta}^v}\right]. \quad (36)\end{aligned}$$

Taking the ratio of the overall mean BTDs gives Eq. (9).

Setting $\rho^v = 0$, we have that

$$G_v^{GPS,L} = \frac{\langle\tilde{\boldsymbol{\rho}}^v,\mathbf{1}-\bar{\boldsymbol{s}}^v\rangle_{\boldsymbol{\Delta}^v}}{s^v\langle\boldsymbol{\delta}^v,\tilde{\boldsymbol{\rho}}^v\rangle + \langle\tilde{\boldsymbol{g}}',\tilde{\boldsymbol{\rho}}^v\rangle_{\boldsymbol{\Delta}^v}}.$$

If we further have $\rho^v \to 0, \forall v \in \mathcal{V}$, then $\tilde{\boldsymbol{g}}' \to \boldsymbol{0}$ and

$$\begin{aligned}G_v^{GPS,L} &= \frac{\langle\tilde{\boldsymbol{\rho}}^v,\mathbf{1}-\bar{\boldsymbol{s}}^v\rangle_{\boldsymbol{\Delta}^v}}{s^v\langle\boldsymbol{\delta}^v,\tilde{\boldsymbol{\rho}}^v\rangle} \\ &= \frac{\langle\boldsymbol{\delta}^v,\tilde{\boldsymbol{\rho}}^v\rangle - (1-s^v)\langle\boldsymbol{\delta}^v,\tilde{\boldsymbol{\rho}}^v\rangle}{s^v\langle\boldsymbol{\delta}^v,\tilde{\boldsymbol{\rho}}^v\rangle} = 1\end{aligned}$$

When $\rho^v \to \infty, \forall v \in \mathcal{V}$, all the terms without $\rho^v$ vanishes and the gain becomes

$$G_v^{GPS,H} \to \frac{\|\tilde{\boldsymbol{\rho}}^v\|_{\boldsymbol{\Delta}^v}^2 - \|\tilde{\boldsymbol{\rho}}^v\|_{\boldsymbol{\Delta}^v\boldsymbol{S}^v}^2}{\langle\tilde{\boldsymbol{g}},\tilde{\boldsymbol{\rho}}^v\rangle_{\boldsymbol{\Delta}^v}}.$$

Note that even we have $\rho^v \to \infty$, we cannot guarantee that $\rho_b^v \to \infty, \forall b \in \mathcal{B}$ if for some $b \in \mathcal{B}, \tilde{\rho}_b^v = 0$. Thus $\|\tilde{\boldsymbol{\rho}}^v\|_{\boldsymbol{S}^v}^2$ might not approach 0.

*D. Proof of Corollary 3*

From Theorem 1, one can express the overall weighted BTD gain over SS under heavy load as

$$\begin{aligned}G_{\text{all}}^{SS,H} &= \frac{\sum_v s^v\|\tilde{\boldsymbol{\rho}}^v\|_2^2}{\sum_v s^v\langle\tilde{\boldsymbol{g}},\tilde{\boldsymbol{\rho}}^v\rangle} = \frac{\sum_v \sum_b s^v(\tilde{\rho}_b^v)^2}{\sum_v s^v \sum_b(\sum_{v'} s^{v'}\tilde{\rho}_b^{v'})\tilde{\rho}_b^v} \\ &= \frac{\sum_b \sum_v s^v(\tilde{\rho}_b^v)^2}{\sum_b(\sum_{v'} s^{v'}\tilde{\rho}_b^{v'})(\sum_v s^v\tilde{\rho}_b^v)} = \frac{\sum_b \sum_v s^v(\tilde{\rho}_b^v)^2}{\sum_b(\sum_v s^v\tilde{\rho}_b^v)^2}.\end{aligned}$$

According to Jensen's inequality, we have $\forall b \in \mathcal{B}$, $\sum_v s^v(\tilde{\rho}_b^v)^2 \geq (\sum_v s^v\tilde{\rho}_b^v)^2$. Thus $G_{\text{all}}^{SS} \geq 1$.

Similarly, for GPS, we have

$$\begin{aligned}G_{\text{all}}^{GPS,H} &= \frac{\sum_v \sum_b s^v(\tilde{\rho}_b^v)^2\left(1-\sum_{v'\neq v} s^{v'}e^{-\rho_b^{v'}}\right)}{\sum_v \sum_b s^v\tilde{\rho}_b^v \sum_{v'} s^{v'}\tilde{\rho}_b^{v'}} \\ &= \frac{\sum_b \sum_v s^v(\tilde{\rho}_b^v)^2 - \sum_b \sum_v s^v(\tilde{\rho}_b^v)^2\sum_{v'\neq v}s^{v'}e^{-\rho_b^{v'}}}{\sum_b(\sum_v s^v\tilde{\rho}_b^v)^2}.\end{aligned}$$

As $\rho^v \to \infty$, for each base station $b \in \mathcal{B}$, if $\rho_b^v \to \infty, e^{-\rho_b^v} \to 0$, otherwise $\tilde{\rho}_b^v = \frac{\rho_b^v}{\rho^v} \to 0$. Based on such observation, let us define a set of slices at each base station $b$, whose local loads approach infinity, $\mathcal{V}_b^{\inf} \triangleq \{v \in \mathcal{V} : \rho_b^v \to \infty\}$. Then the above equation can be rewritten as:

$$G_{\text{all}}^{GPS,H} = \frac{\sum_b \left( \sum_{v \in \mathcal{V}_b^{\inf}} s^v (\tilde{\rho}_b^v)^2 (1 - \sum_{v' \notin \mathcal{V}_b^{\inf}} s^{v'} e^{-\rho_b^{v'}}) \right)}{\sum_b (\sum_{v \in \mathcal{V}_b^{\inf}} s^v \tilde{\rho}_b^v)^2}$$

$$= \frac{\sum_b \left( (1 - \sum_{v' \notin \mathcal{V}_b^{\inf}} s^{v'} e^{-\rho_b^{v'}}) (\sum_{v \in \mathcal{V}_b^{\inf}} s^v) \sum_{v \in \mathcal{V}_b^{\inf}} \tilde{s}^v (\tilde{\rho}_b^v)^2 \right)}{\sum_b (\sum_{v \in \mathcal{V}_b^{\inf}} s^v)^2 (\sum_{v \in \mathcal{V}_b^{\inf}} \tilde{s}^v \tilde{\rho}_b^v)^2},$$

where for $v \in \mathcal{V}_b^{\inf}$, $\tilde{s}^v \triangleq \frac{s^v}{\sum_{v \in \mathcal{V}_b^{\inf}} s^v}$. Therefore $\sum_{v \in \mathcal{V}_b^{\inf}} \tilde{s}^v = 1$. Now for each base station $b \in \mathcal{B}$, we have

$$\frac{(1 - \sum_{v' \notin \mathcal{V}_b^{\inf}} s^{v'} e^{-\rho_b^{v'}})(\sum_{v \in \mathcal{V}_b^{\inf}} s^v) \sum_{v \in \mathcal{V}_b^{\inf}} \tilde{s}^v (\tilde{\rho}_b^v)^2}{(\sum_{v \in \mathcal{V}_b^{\inf}} s^v)^2 (\sum_{v \in \mathcal{V}_b^{\inf}} \tilde{s}^v \tilde{\rho}_b^v)^2}$$

$$= \frac{(1 - \sum_{v \notin \mathcal{V}_b^{\inf}} s_v e^{-\rho_b^v}) \sum_{v \in \mathcal{V}_b^{\inf}} \tilde{s}^v (\tilde{\rho}_b^v)^2}{(\sum_{v \in \mathcal{V}_b^{\inf}} s^v)(\sum_{v \in \mathcal{V}_b^{\inf}} \tilde{s}^v \tilde{\rho}_b^v)^2}$$

$$\geq \frac{(1 - \sum_{v \notin \mathcal{V}_b^{\inf}} s_v) \sum_{v \in \mathcal{V}_b^{\inf}} \tilde{s}^v (\tilde{\rho}_b^v)^2}{(\sum_{v \in \mathcal{V}_b^{\inf}} s^v)(\sum_{v \in \mathcal{V}_b^{\inf}} \tilde{s}^v \tilde{\rho}_b^v)^2} = \frac{\sum_{v \in \mathcal{V}_b^{\inf}} \tilde{s}^v (\tilde{\rho}_b^v)^2}{(\sum_{v \in \mathcal{V}_b^{\inf}} \tilde{s}^v \tilde{\rho}_b^v)^2},$$

where the last equality holds true because $\sum_v s^v = 1$. Then by Jensen's inequality, for all $b \in \mathcal{B}$, the above ratio is no less than 1, thus $G_{\text{all}}^{GPS,H} \geq 1$.

*E. Proof of Theorem 5*

Under the saturated regime, BTD constraint of each $v \in \mathcal{V}$ must be binding because we are blocking traffics. Thus we have:

$$x_v = \frac{f_v(\tilde{\rho}^v; \tilde{\rho}^{-v})}{s^v(\tilde{d}_v - 1)}, \tag{37}$$

Also assumption 3 guarantees that for all $v \in \mathcal{V}$, $M^v \tilde{\rho}^v \preceq x_v \mathbf{1}$ is not binding then the $\mathbf{AC}_v$ can be reformulated as:

$$\min_{\tilde{\rho}^v \in \Gamma^v} s^v f_v(\tilde{\rho}^v; \tilde{\rho}^{-v}), \tag{38}$$

A constant factor $s^v$ in the objective function has no impact to the minimizer. We can show that the optimality condition of Problem (28) is the same as each slice $v$ optimizing its own problem given by Eq. (38). Dividing the objective function by 2, the Lagrangian of Problem (28) is:

$$L(\tilde{\rho}; \zeta, \chi) = \frac{1}{2}(\|\sum_v s^v \tilde{\rho}^v\|_2^2 + \sum_v (s^v)^2 \|\tilde{\rho}^v\|_2^2)$$

$$+ \sum_{v \in \mathcal{V}} \zeta_v(\langle \mathbf{1}, \tilde{\rho}^v \rangle - 1) - \sum_{v \in \mathcal{V}} (\chi^v)^T M_v \tilde{\rho}^v, \tag{39}$$

where $\tilde{\rho} \triangleq (\tilde{\rho}^v : v \in \mathcal{V})$, dual variables $\zeta \triangleq (\zeta_v : v \in \mathcal{V})$, and $\chi \triangleq (\chi^v : v \in \mathcal{V})$. According to the KKT condition, the solution $\tilde{\rho}^*$ must be such that, for all $v \in \mathcal{V}$:

$$\nabla_{\tilde{\rho}^v} L(\rho^*; \zeta^*, \chi^*) = (s^v)^2 \tilde{\rho}^{v,*} + s^v \tilde{g} + \zeta_v^* \mathbf{1} - (M^v)^T \chi^{v,*}$$
$$= \mathbf{0}, \tag{40}$$

and $\chi^{v,*} \succeq \mathbf{0}, \tilde{\rho}^{v,*} \in \Gamma^v$. The Lagrangian of Problem (38) is:

$$L_v(\tilde{\rho}^v; \zeta_v, \chi^v) = \langle \tilde{g}, s^v \tilde{\rho}^v \rangle$$
$$+ \zeta_v(\langle \mathbf{1}, \tilde{\rho}^v \rangle - 1) - (\chi^v)^T M_v \tilde{\rho}^v. \tag{41}$$

If slice $v$'s relative load $\tilde{\rho}^{v,*}$ optimizes Problem (38) given other slices' $\tilde{\rho}^{-v,*}$, following KKT condition should be met:

$$\nabla_{\tilde{\rho}^{v,*}} L_v(\tilde{\rho}^{v,*}; \zeta_v^*, \chi^{v,*}) = (s^v)^2 \tilde{\rho}^{v,*} + s^v \tilde{g}$$
$$+ \zeta_v^* \mathbf{1} - (M_v^{-1})^T \chi^{v,*} = \mathbf{0}, \tag{42}$$

and $\chi^{v,*} \succeq \mathbf{0}, \tilde{\rho}^{v,*} \in \Gamma^v$, which is exactly the same as the KKT condition of Problem (28). Therefore, Problem (28) is solved at the Nash equilibrium. Moreover, we could compute the total carried load of slice $v$ by setting

$$f_v(\tilde{\rho}^{v,*}; \tilde{\rho}^{-v,*}) = s^v(\tilde{d}_v - 1)/\rho^{v,*},$$

which gives us $\rho^{v,*} = \frac{s^v(\tilde{d}_v - 1)}{\langle \tilde{g}^*, \tilde{\rho}^{v,*} \rangle}$.

*F. Proof of Theorem 6*

Under the saturated regime, BTD constraint of each $v \in \mathcal{V}$ must be binding. Thus we have:

$$\rho^v = \frac{s^v \tilde{d}_v - 1}{\|\tilde{\rho}^v\|_2^2}.$$

Moreover, the constraint $a^v \preceq 1$ should be satisfied with strict inequality, thus the optimal policy $\tilde{\rho}^{v,SS,*}$ under SS is given by:

$$\min_{\tilde{\rho}^v} \{ \|\tilde{\rho}^v\|_2^2 \mid M^v \tilde{\rho}^v \succeq 0, \quad \langle \mathbf{1}, \tilde{\rho}^v \rangle = 1 \}.$$

Then the optimal load is obtained by plugging the result in the BTD constraint.

*G. Proof of Corollary 4*

Under SS, it is obvious that the optimal relative load distribution of slice $v$ is

$$\tilde{\rho}^{v,SS,*} = \frac{1}{B} \mathbf{1}$$

Plugging it in the BTD constraint one can get $\rho^{v,SS,*} = B(s^v \tilde{d}_v - 1)$, thus $\boldsymbol{\rho}^{v,SS,*} = (s^v \tilde{d}_v - 1)\mathbf{1}$.

Dividing the objective function by $s^v$ and discarding the routing constraints, the Lagrangian of Eq. (38) is:

$$\langle \tilde{g}^v, \tilde{\rho}^v \rangle + \nu(\langle \mathbf{1}, \tilde{\rho}^v \rangle - 1),$$

where $\nu$ is the dual variable. Solving its KKT condition we have:

$$\tilde{\rho}^{v,*} = -\frac{1}{2s^v}(\nu \mathbf{1} + \sum_{v' \neq v} s_{v'} \tilde{\rho}^{v'}). \tag{43}$$

Substituting in $\langle \mathbf{1}, \tilde{\rho}^{v,*} \rangle = 1$ we have $\nu = -\frac{s^v + 1}{B}$.

When $\frac{1}{B}\mathbf{1} \in \Gamma^v, \forall v \in \mathcal{V}$, Eq. (43) implies that if all other slices $v' \neq v$ pick their relative loads as $\frac{1}{B}\mathbf{1}$, then $(\sum_{v' \neq v} s^{v'} \tilde{\rho}^{v'}) \parallel \mathbf{1}$, meaning that this is the Nash equilibrium of the game. Note that since $\tilde{\rho}^{v,*}$ is feasible and optimal for a relaxed feasible set, it will still be optimal if we put back the routing constraints. Thus we have that: $\langle \tilde{g}^*, \tilde{\rho}^{v,*} \rangle = \frac{1}{B}$. Then the carried load gain is obtained by plugging the result in Eq. (30).